\documentclass[aps,prd,showpacs,preprint]{revtex4-1}

\usepackage{amsmath,amsfonts,amssymb}
\usepackage{graphicx}
\usepackage{amsmath}

\begin{document}

\title{Cosmological model with decaying vacuum energy law from principles of quantum mechanics} 
\author{Marek Szyd{\l}owski}
\email{marek.szydlowski@uj.edu.pl}
\affiliation{Astronomical Observatory, Jagiellonian University, Orla 171, 30-244 Krak{\'o}w, Poland}
\affiliation{Marc Kac Complex Systems Research Centre, Jagiellonian University,
Reymonta 4, 30-059 Krak{\'o}w, Poland}

\date{\today}

\begin{abstract}
We construct the cosmological model to explain the cosmological constant problem. We built the extension of the standard cosmological model $\Lambda$CDM by consideration of decaying vacuum energy represented by the running cosmological term. From the principles of quantum mechanics one can find that in the long term behavior survival probability of unstable states is a decreasing function of the cosmological time and has the inverse power-like form. This implies that cosmological constant $\rho_{\text{vac}} = \Lambda(t) = \Lambda_{\text{bare}} + \frac{\alpha}{t^2}$ where $\Lambda_{\text{bare}}$ and $\alpha$ are constants. We investigate the dynamics of this model using dynamical system methods due to a link to the $\Lambda(H)$ cosmologies. We have found the exact solution for the scale factor as well as the indicators of its variability like the deceleration parameter and the jerk. From the calculation of the jerk we obtain a simple test of the decaying vacuum in the FRW universe. Using astronomical data (SNIa, $H(z)$, CMB, BAO) we have estimated the model parameters and compared this model with the $\Lambda$CDM model. Our statistical results indicate that the decaying vacuum model is a little worse than the $\Lambda$CDM model. But the decaying vacuum cosmological model explains the small value of the cosmological constant today.
\end{abstract}

\pacs{98.80.Bp, 98.80.Cq, 11.25.-w}

\maketitle

\section{Introduction}

The standard cosmological model ($\Lambda$CDM model) is a very good description of the evolution of the Universe in the current epoch. However it is only an effective description in the terms of dark matter and dark energy of unknown nature. The natural way of explanation why our Universe's expansion is accelerating is to extend the CDM model with the constant cosmological term. On the other hand it seems to be natural to interpret the cosmological constant as the vacuum energy \cite{Weinberg:1988cp}. Although such an explanation brings the answer qualitatively the calculations of the value of the cosmological constant term from the quantum field theory and its value required for solving the Universe accelerated expansion conundrum give rise to discrepancy, unknown in the history of physics, of more than 100 orders of magnitude. Therefore the problem of cosmological constant is crucial for consistency of the standard cosmological model. 

In the literature different approaches have been taken to explain why the value of the cosmological constant is so small today. The most popular approach is description of the dark sector of the Universe in terms of the self-interacting scalar field with a potential \cite{Alam:2003rw}. Alternatively in this paper we develop the Alcaniz and Lima approach in which cosmology with the decay vacuum energy is interpreted in the framework of the $\Lambda(t)$CDM cosmology \cite{Alcaniz:2005dg,Wang:2004cp,Graef:2013iia}. Alcaniz and Lima \cite{Alcaniz:2005dg} pointed out that the critique of these models concentrates on lack of the theoretical model which can specify the phenomenological form of the relation describing how $\Lambda$ depends on time. In this paper we are going to address the theoretical foundation of this approach.

The discussion of false vacuum in the context of cosmology has been started by Coleman \cite{Coleman:1977py,Callan:1977pt}. The decaying of false vacuum in the context of inflation has been investigated by Krauss and Dent \cite{Krauss:2007rx}. The properties of unstable false vacuum can be investigated from the point of view of the quantum theory of unstable states \cite{Urbanowski:2012pka,Urbanowski:2013tfa} and some universal behavior of survival probability of vacuum states can be discovered. 

From the quantum theory the amplitude $A(t)$ and the decay law $\mathcal{P}_{M}(t)$ of unstable state $| M \rangle$ are determined by energy density distribution function $\omega(E)$ for the system in this state. 
\begin{equation}
A(t) = \int_{E_{\text{min}}}^{\infty} \omega(E) \exp\left( - \frac{i}{\hbar} Et \right) dt.
\end{equation}
where $\omega(E) \ge 0$ and $\omega(E) = 0$ for $E < E_{\text{min}}$.

The last condition and the Paley-Wiener theorem give us $|A(t)| \ge A_1 \exp(-A_2 t^q)$ for $t \to \infty$. In consequence the vacuum unstable states decay following the power-law relation after exponential for $t \gg T$ where $T$ is the characteristic time of a regime switching. The asymptotically we obtain universal relation
\begin{equation}
E_{0}^{\text{false}}(t) = E_{0}^{\text{false}}(t) \pm \frac{\alpha^2}{t^2} \pm \cdots, \quad \text{for } t \gg T.
\end{equation}

At the late time survival probability assumes the form of the power-law relation. We demonstrate that then the instantaneous energy of the vacuum states has universal asymptotics $\rho_{\text{vac}} = \rho_{\text{vac}}(t) \cong \rho_{0}^{\text{true}} + \frac{\alpha}{t^2} \neq \rho_{0}^{\text{false}}$. Using this parameterization of the false vacuum will study the evolution of the Friedman-Robertson-Walker (FRW) universe filled with dust matter and dark energy running in the cosmological time following the rule
\begin{equation}\label{eq:2}
\rho_{\text{vac}} = \Lambda(t) = \Lambda_{\text{bare}} + \frac{\alpha}{t^2}
\end{equation}
where we use the natural system of units $8\pi G = c =1$.

The parametrization (\ref{eq:2}) is only a leading term at the late time of evolution. In general from principles of quantum mechanics it assumes more general form in the form of power law series \cite{Urbanowski:2012pka}
\begin{equation}
\Lambda(t)=\Lambda + \frac{\alpha}{t^2} + \frac{\beta}{t^4} + \frac{\gamma}{t^6} + \cdots
\end{equation}

The relation (\ref{eq:2}) was proposed ad-hoc in the context of cosmological constant problem by Lopez and Nanopoulos \cite{Lopez:1995eb}. They showed that starting at the Planck epoch $\Lambda_{\text{Pl}} \sim M_{\text{Pl}}^{2}$ the Universe reaches today epoch with $\Lambda_{0} \sim 10^{120} M_{\text{Pl}^{2}}$. Another similar result was obtained by Lima et al. \cite{Lima:2015kda}.

The main aim of this paper is to explain the cosmological constant problem. We propose the theoretical justification of choosing the varying $\Lambda$ during the cosmic evolution parametrized as (\ref{eq:2}). The form of this parametrization is motivated by quantum mechanics principles. The key point of this approach is the connection of decaying vacuum with measurement of a running jerk. The dimensionless parameter $\alpha$ is estimated from astronomical data.

\section{The exact solution for the scale factor in the FRW decaying vacuum cosmology}

Throughout this paper we consider a flat homogeneous and isotropic universe in which a source of gravity is the perfect fluid with energy density $\rho(t)$ and pressure $p(t)$. We introduce the energy momentum tensor
\begin{equation}\label{eq:4}
T^{\alpha \beta} = T_{\text{m}}^{\alpha \beta} + \Lambda(t) g^{\alpha \beta}
\end{equation}
where the first component describes dust and the second one the time-dependent $\Lambda$ term. 

The energy-momentum conservation condition $T^{\alpha \beta}_{\, \, ; \beta} = 0$ gives rise to the condition
\begin{equation}\label{eq:5}
\dot{\rho}_{\text{m}} + 3 \frac{\dot{a}}{a} \rho_{\text{m}} = - \dot{\rho}_{\text{vac}}
\end{equation}
where $a(t)$ is the scale factor, $\rho_{\text{m}}$ and $\rho_{\text{vac}}$ are the energy densities of cold dark matter and vacuum, respectively, a dot means the differentiation with respect to the cosmological time~$t$, $\dot{\rho}_{\text{vac}} = \frac{d\Lambda}{dt}$ and the equation of state for vacuum is $\rho_{\text{vac}} = - p_{\text{vac}}$ where $p_{\text{vac}}$ is pressure.

The Einstein equations reduce to the acceleration equation
\begin{equation}
3 \frac{\ddot{a}}{a} = - \frac{1}{2} \rho_{\text{m}} + \Lambda(t)
\end{equation}
where $\Lambda(t)$ is given by formula (\ref{eq:2}). Equation (\ref{eq:5}) can be interpreted formally within the framework of interacting cosmology
\begin{equation}
\dot{\rho}_{\text{m}} + 3 \frac{\dot{a}}{a} \rho_{\text{m}} = \Gamma(t)
\end{equation}
where $\Gamma = - \dot{\rho}_{\text{vac}} = - \dot{\Lambda}$ is the energy exchange term. This term is called a source term for the particle creation process \cite{Bessada:2013maa}. If $\Gamma(t)$ is positive then the vacuum energy density must be decaying. Because of parametrization (\ref{eq:2}) we have $\Gamma(t) = 2 \alpha t^{-3}$, which is positive for the positive $\alpha$.

For our aim it would be useful to rewrite the vacuum dark energy into the new parametrization. The idea of Alcaniz and Lima \cite{Alcaniz:2005dg} is that the decay vacuum should modify the law of evolution of the cold dark matter ($\rho_{\text{m}} \propto a^{-3}$) in such a way that matter dilutes more slowly like
\begin{equation}\label{eq:6}
\rho_{\text{m}} = \rho_{\text{m},0} a^{-3+\varepsilon}
\end{equation}
where $\varepsilon$ is a small positive constant in $(0,1)$.

Thus, we substitute equation (\ref{eq:6}) into the continuity condition (\ref{eq:5}) and obtain
\begin{equation}\label{eq:7}
\rho_{\text{vac}} = \rho_{\text{vac},0} + \frac{\varepsilon}{3-\varepsilon} \rho_{\text{m}} a^{-3+\varepsilon}.
\end{equation}
The constant $\rho_{\text{vac},0}$ is called ``the ground state value of the vacuum'' \cite{Wang:2004cp}. We interpret $\rho_{\text{vac},0}$ as $\Lambda_{\text{bare}}$ -- the constant contribution to the vacuum energy parametrization. And we interpret the last term in equation (\ref{eq:7}) 
as $\alpha/t^2$ following the parametrization (\ref{eq:2}). Hence after elementary calculations we obtain that
\begin{equation}\label{eq:8}
\frac{\alpha}{t^2} = \frac{3-\varepsilon}{2} \alpha H^2
\end{equation}
or
\begin{equation}\label{eq:8b}
\Lambda = \Lambda(H) = \Lambda_{\text{bare}} + 3 \beta H^2
\end{equation}
where $\beta = \frac{3-\varepsilon}{2} \alpha$. We assume that $0 < \varepsilon < 1$, and  

As a consequence we obtain the phenomenological quadratic parametrization of $\Lambda$ through the Hubble parameter $H$ \cite{Lima:1995kd}. The same form of parameterization can be also obtained after replacing the cosmological time $t$ by the Hubble scale $t_{H} = 1/H$.

Now it would be useful to rewrite the acceleration equation to the new form (where constant contribution $\Lambda_{\text{bare}}$ is denoted as $\Lambda$ in further analysis)
\begin{equation}\label{eq:9}
\frac{dH}{dt} = \dot{H} = \frac{\Lambda}{2} - \delta H^2
\end{equation}
where $\delta$ is constant defined as
\begin{equation}\label{eq:10}
\delta = \frac{3}{2}(1-\beta).
\end{equation}
It is convenient for our purposes to introduce a new time variable into equation (\ref{eq:9}). Let assume that there is the inverse function of $t=t(a)$ for the cosmic evolution. Then
\begin{equation}\label{eq:11}
\frac{dH}{dt} = \frac{dH}{da} \frac{da}{dt} = H a \frac{dH}{da} = \frac{\Lambda}{2} - \delta H^2
\end{equation}
or
\begin{equation}\label{eq:12}
\frac{dH}{da} = H' = \frac{1}{aH} \left( \frac{\Lambda}{2} - \delta H^2 \right)
\end{equation}
where a prime denotes a derivative with respect to the scale factor $a$. Note that if $\beta=0$ or $\delta = 3/2$ we obtain a corresponding relation for the $\Lambda$CDM model.

Now one can check that the solution of equation (\ref{eq:12}) is given in the form
\begin{equation}\label{eq:13}
H^2(a) = \left( H_{0}^{2} - \frac{\Lambda}{2\delta} \right) a^{-2\delta} + \frac{\Lambda}{2\delta}
\end{equation}
where $H=H_0$ for $a=a_0 =1$ and all quantities with a subscript ``0'' are evaluated at the present epoch. 

In the context of cosmography it would be useful to rewrite relation (\ref{eq:13}) to the new form in which appears dimensionless density parameters for the cold dark matter and the constant contribution to the dynamical dark energy, $\Lambda$, parameterized following formula (\ref{eq:2})
\begin{equation}\label{eq:14}
\Omega_{\text{m},0} = \frac{\rho_{\text{m},0}}{3H_{0}^{2}}, \qquad \Omega_{\Lambda,0} = \frac{\Lambda}{3H_{0}^{2}}.
\end{equation}
Then we obtain the relation $H(z)$ for the extended $\Lambda$CDM model with decaying vacuum dark energy in the form
\begin{equation}\label{eq:15}
\frac{H(z)}{H_{0}} = \sqrt{\frac{\Omega_{\text{m},0}}{\Omega_{\text{m},0}+\Omega_{\Lambda_{0}}} (1+z)^{3(\Omega_{\text{m},0} + \Omega_{\Lambda,0})} + \frac{\Omega_{\Lambda,0}}{\Omega_{\text{m},0}+\Omega_{\Lambda,0}}}
\end{equation}
where the redshift $z \equiv (1+a)^{-1}$. If $\Omega_{\text{m},0}+\Omega_{\Lambda,0}=1$ we obtain the corresponding relation for the $\Lambda$CDM model. Therefore, the effects of decaying vacuum are manifested by violation of the condition $\Omega_{\text{m},0}+\Omega_{\Lambda,0} = 1$.

The parameter $\beta$ in the parameterization of $\Lambda(H)$ (\ref{eq:8b}) was constrained by using the observations of cosmic star formation and the age of the Universe as $3\beta \lesssim 0.3$ \cite{Bessada:2013maa}. In the next section we will search for more stringent upper limit for this parameter.

Let us consider an exact solution for the function (\ref{eq:13}). The scale factor $a$ satisfies a first order differential equation
\begin{equation}\label{eq:16}
\dot{a}^{2} = H_{0}^{2} \left( \frac{\Omega_{\text{m},0}}{\Omega_{\text{m},0}+\Omega_{\Lambda_{0}}} a^{-3(\Omega_{\text{m},0} + \Omega_{\Lambda,0})+2} + \frac{\Omega_{\Lambda,0}}{\Omega_{\text{m},0}+\Omega_{\Lambda_{0}}} a^{2} \right).
\end{equation}
Equation (\ref{eq:16}) can be integrated after a simple substitution
\begin{equation}\label{eq:17}
a \to x \colon x^2 = a^{3(\Omega_{\text{m},0} + \Omega_{\Lambda,0})}
\end{equation}
and the time parametrization
\begin{equation}\label{eq:17b}
t \to \tau \colon \tau = \frac{3}{2} |H_0| \sqrt{\Omega_{\Lambda,0}} \, t.
\end{equation}
After solving the equation for $x(\tau)$ and returning to original variables we obtain the following expression for the scale factor as a function of the cosmological time
\begin{equation}\label{eq:18} 
a(t) = \left( \frac{\Omega_{\text{m},0}}{\Omega_{\Lambda,0}} \right)^{\frac{1}{3(\Omega_{\text{m},0} + \Omega_{\Lambda,0})}} \left( \sinh \frac{3}{2} \sqrt{\Omega_{\Lambda,0}} \, |H_0| t \right)^{\frac{2}{3(\Omega_{\text{m},0} + \Omega_{\Lambda,0})}}
\end{equation}
Formula (\ref{eq:18}) gives us the possibility to calculate the cosmological scalars which control the variability of the cosmic evolution
\begin{align}
H &= \frac{1}{a} \frac{da}{dt} \nonumber \\
q &= -a \left( \frac{da}{dt} \right)^{-2} \frac{d^2 a}{dt^2} \nonumber \\
Q &= a^2 \left( \frac{da}{dt} \right)^{-3} \frac{d^3 a}{dt^3} \\
Q &= a^3 \left( \frac{da}{dt} \right)^{-4} \frac{d^4 a}{dt^4} \nonumber \\
Q &= a^4 \left( \frac{da}{dt} \right)^{-5} \frac{d^5 a}{dt^5} \nonumber
\end{align}
These scalars are called the Hubble parameter, deceleration parameter, jerk, snap and crackle, respectively. They are constrained by some algebraic relation \cite{Dunajski:2008tg}.

Let us calculate the cosmic jerk related with a third order time derivative of the scale factor
\begin{equation}\label{eq:20}
Q = \frac{(\Delta - 1)(\Delta - 2)}{\Delta^2} + \frac{(3\Delta - 2)}{\Delta^2} \tanh^{2} \left( \frac{3}{2} |H_0| \sqrt{\Omega_{\Lambda,0}} \, t \right)
\end{equation}
where $\Delta = \frac{2}{3} (\Omega_{\text{m},0} + \Omega_{\Lambda,0})$.

If $\Delta = \frac{2}{3}$ we have the case of the $\Lambda$CDM model for which $Q=1$ and if $\Delta \neq \frac{2}{3}$ or $\Omega_{\text{m},0} + \Omega_{\Lambda,0}$ then $Q=Q(t)$. Therefore if we find that the jerk is different from one it will be \textit{experimentum crucis\/} for excluding the $\Lambda$CDM model. 
Then as the alternative the cosmological model with decaying vacuum should be treated seriously. In this model the jerk is time dependent and now different from one as the function of sum of $\Omega_{\text{m},0}$ and $\Omega_{\Lambda,0}$ which is different from one.

\section{Dynamic of the model on the phase plane}

For deeper analysis of the dynamics of the model let us formulate this model as a dynamical system. 
\begin{align}\label{eq:23}
\dot{H} &= - H^2 - \frac{1}{6} \rho_{\text{m}} + \frac{1}{3}(\Lambda + \Lambda(H)) \\ \label{eq:24}
\dot{\rho}_{\text{m}} &= -3H\rho_{\text{m}} - \frac{d\Lambda}{dH} \left[ - H^2 - \frac{1}{6} \rho_{\text{m}} + \frac{1}{3}(\Lambda + \Lambda(H)) \right].
\end{align}
Dynamical system (\ref{eq:23})-(\ref{eq:24}) assumes the form of two-dimensional autonomous dynamical system. In the special case of $\Lambda(H)=\alpha H^2$ and
\begin{align}\label{eq:25}
\dot{H} &= \left( \frac{\alpha}{3} - 1 \right) H^2 - \frac{1}{6} \rho_{\text{m}} + \frac{1}{3}\Lambda  \\ \label{eq:26}
\dot{\rho}_{\text{m}} &= \left( \frac{\alpha}{3} - 3 \right) H\rho_{\text{m}} + 2 \alpha \left( \frac{\alpha}{3} -1 \right) H^3 - \frac{2}{3} \alpha H.
\end{align}
It could be useful to rewrite the above system in two state variables and new time variable
\begin{equation}
x = H^2, \qquad y = \rho_{\text{m}}, \qquad \tau = \ln a.
\end{equation}
This system possesses the critical point 
\begin{equation}
x_0 = \frac{\Lambda}{\alpha - 3}, \qquad y_0 = 0.
\end{equation}
Now we translate the critical point at the origin of the coordinate system as follow
\begin{equation}
X = x - x_0 \qquad Y = y - y_0.
\end{equation}
Finally the dynamical system possesses the form of linear system 
\begin{align} \label{eq:32}
X'(\tau ) &= \text{  }\frac{2(\alpha -3)}{3}X(\tau )-\frac{1}{3}Y(\tau ) \\
\label{eq:33}
Y'(\tau ) &= -\frac{2(\alpha -3)\alpha  }{3}X(\tau )+\frac{(\alpha -9)}{3}Y(\tau )
\end{align}

\begin{figure}
\centering
\includegraphics[width=0.8\textwidth]{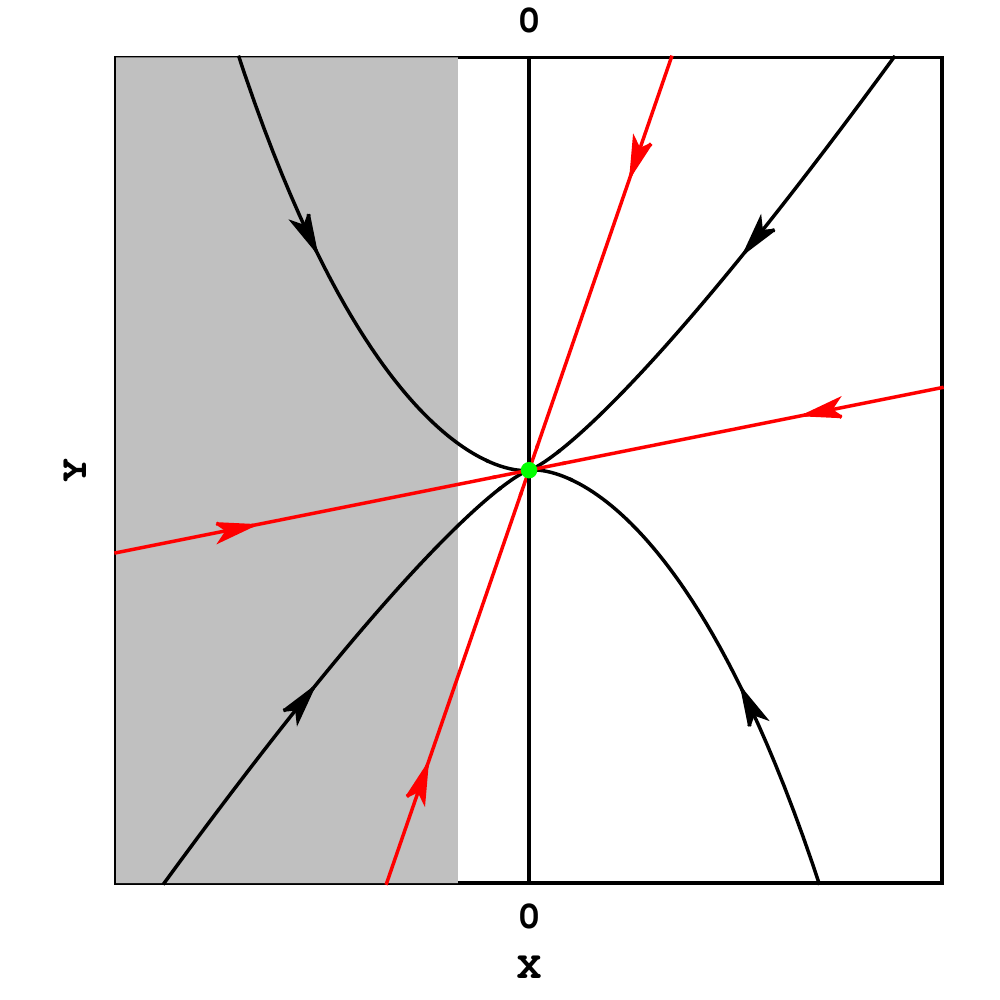}
\caption{The phase portrait of system (\ref{eq:32})-(\ref{eq:33}) with the positive cosmological constant $\Lambda$ and $\Lambda(H) = \alpha H^2$. The critical point at the origin represents the de Sitter solution.}
\label{fig:1}
\end{figure}

The phase portrait for system (\ref{eq:32})-(\ref{eq:33}) with the positive cosmological constant is presented in Fig.~\ref{fig:1}. The system possesses the critical point which is stable node. It is representing the de Sitter solution with the cosmological constant $\Lambda$. The physical region is determined by condition $x = H^2 \le 0$ and $y \ge 0$. Therefore the physical trajectories should lie in the first quarter of $(x,y)$-coordinate systems. Note that if $\Lambda$ is negative the unique critical point is situated in the non-physical region. 

The solution of the system in the parametric form is
\begin{align}
X(\tau ) &= \frac{1}{3(\alpha - 1)} \left[\left(2 \alpha  e^{(\alpha -3) \tau } +(\alpha -3)e^{-2 \tau }\right) C_1+\left(e^{-2 \tau
}-e^{(\alpha -3) \tau }\right) C_2\right] \\
Y(\tau ) &= \frac{1 }{3 (\alpha -1)}\left[\left( e^{(\alpha -3) \tau } -e^{-2 \tau }\right)2\alpha (\alpha -3) C_1+ \left(2 \alpha  e^{-2 \tau
}+(\alpha -3)e^{(\alpha -3) \tau }\right) C_2\right]
\end{align}
Note that there are four special solution lying along two the straight lines 
\begin{equation}
Y(\tau) = \beta X(\tau)
\end{equation}
where the parameter $\beta$ is
\begin{equation}
\beta = 3-\alpha
\end{equation}
or
\begin{equation}
\beta = 2\alpha.
\end{equation}
The line $x=0$ separates the cosmological models with a singularity from singularity-free models. The latter are starting from the boundary of non-physical region and going toward the de Sitter global attractor. The former are starting from singularities $(H=\infty, \rho_{\text{m}}=\infty)$ of the Einstein-de Sitter or Milne type.

For full analysis of the system dynamics it is necessary to know how the system trajectories behave at the infinity. For this aim we will construct the global phase portraits. In the case when right-hand sides of a dynamical system are given in a polynomial form of order $n$ the projective coordinates can be introduced. For a two-dimensional dynamical system with state variables $(x,y)$ two projective maps
\begin{align}
\text{I} \colon \qquad z &= \frac{1}{x}, \quad u = \frac{y}{x} \\
\text{II} \colon \qquad w &= \frac{1}{y}, \quad v = \frac{x}{y}.
\end{align}
cover the dynamics. Of course when $u \neq 0$ and $v \neq 0$ these two projective coordinates $(z,u)$ and $(v,w)$ are equivalent. The infinitely distant points of the plane $(x,y)$ in the projective coordinates correspond to a circle $S^1: x^2 + y^2 = \infty$ covered by two straight lines: $z=0$, $-\infty < u < +\infty$ and $w=0$, $-\infty < v < +\infty$. The original dynamical system in state variable $(x,y)$: $\dot{x}= P(x,y)$, $\dot{y}=Q(x,y)$ after transforming to the projective coordinates and reparametrization of the time variable $\tau \to \tau_1 \colon d\tau_1 = x^{n-1} d\tau$ assumes the form
\begin{align}
\dot{z} &= -z P^{*}(z,u), \\
\dot{u} &= Q^{*}(z,u) - u P^{*}(z,u)
\end{align}
where 
\begin{equation}
P^{*}(z,u) = z^{n} P\left( \frac{1}{z}, \frac{u}{z} \right), \qquad
Q^{*}(z,u) = z^{n} Q\left( \frac{1}{z}, \frac{u}{z} \right)
\end{equation}
are polynomials in the projective variable $(z,u)$. 

The analogous procedure can be performed in the map $(v,w)$.

Dynamical system (\ref{eq:32})-(\ref{eq:33}) in the projective coordinates $z(\tau)=\frac{1}{X(\tau)}$, $u(\tau)=\frac{Y(\tau)}{X(\tau)}$ has the following form 
\begin{align} \label{eq:44}
z'(\tau) &=\frac{1}{3}u(\tau)z(\tau)-\frac{2}{3}(\alpha-3)z(\tau) \\
\label{eq:45}
u'(\tau) & =\frac{1}{3}u^2(\tau)-\frac{1}{3}(\alpha+3)u(\tau)-\frac{2}{3}(\alpha-3)\alpha.
\end{align}   
The exact solution of the system (\ref{eq:44})-(\ref{eq:45}) assumes the following form
\begin{align}
z(\tau ) &=\frac{3 (\alpha -1)}{\left[2 \alpha  e^{(\alpha -3) \tau } +(\alpha -3)e^{-2 \tau }\right] C_1+\left(e^{-2 \tau
}-e^{(\alpha -3) \tau }\right) C_2} \\
u(\tau ) &=\frac{\left( e^{(\alpha -3) \tau } -e^{-2 \tau }\right)2\alpha (\alpha -3) C_1+ \left(2 \alpha  e^{-2 \tau
}+(\alpha -3)e^{(\alpha -3) \tau }\right) C_2}{\left[2 \alpha  e^{(\alpha -3) \tau } +(\alpha -3)e^{-2 \tau }\right] C_1+\left(e^{-2 \tau
}-e^{(\alpha -3) \tau }\right) C_2}.
\end{align}
The critical points are $z=0$, $u=3-a$ and $z=0$, $u=2a$. The former is an unstable node and the latter is a saddle.

Dynamical system (\ref{eq:32})-(\ref{eq:33}) in the projective coordinates $w(\tau)=\frac{1}{Y(\tau)}$, $v(\tau)=\frac{X(\tau)}{Y(\tau)}$ has the following form
\begin{align} \label{eq:48}
    w'(\tau) &=\frac{2}{3}(\alpha-3)\alpha w(\tau)v(\tau)-\frac{1}{3}(\alpha-9)w(\tau) \\
\label{eq:49}
    v'(\tau) &=\frac{2}{3}(\alpha-3)\alpha v^2(\tau)+\frac{1}{3}(\alpha+3)v(\tau)-\frac{1}{3}.
\end{align}
The exact solution of the system (\ref{eq:48})-(\ref{eq:49}) assumes the following form
\begin{align}
w(\tau) &=\frac{3 (\alpha -1)}{\left( e^{(\alpha -3) \tau } -e^{-2 \tau }\right)2\alpha (\alpha -3) C_1+ \left[2 \alpha  e^{-2 \tau
}+(\alpha -3)e^{(\alpha -3) \tau }\right] C_2} \\
v(\tau) &=\frac{\left(2 \alpha  e^{(\alpha -3) \tau } +(\alpha -3)e^{-2 \tau }\right) C_1+\left(e^{-2 \tau
}-e^{(\alpha -3) \tau }\right) C_2}{\left( e^{(\alpha -3) \tau } -e^{-2 \tau }\right)2\alpha (\alpha -3) C_1+ \left[2 \alpha  e^{-2 \tau
}+(\alpha -3)e^{(\alpha -3) \tau }\right] C_2}
\end{align}
The critical points are $w=0$, $v=\frac{1}{3-a}$ and $w=0$, $v=\frac{1}{2a}$. The former is an unstable node and the latter is a saddle.

\begin{figure}
\centering
\includegraphics[width=0.8\textwidth]{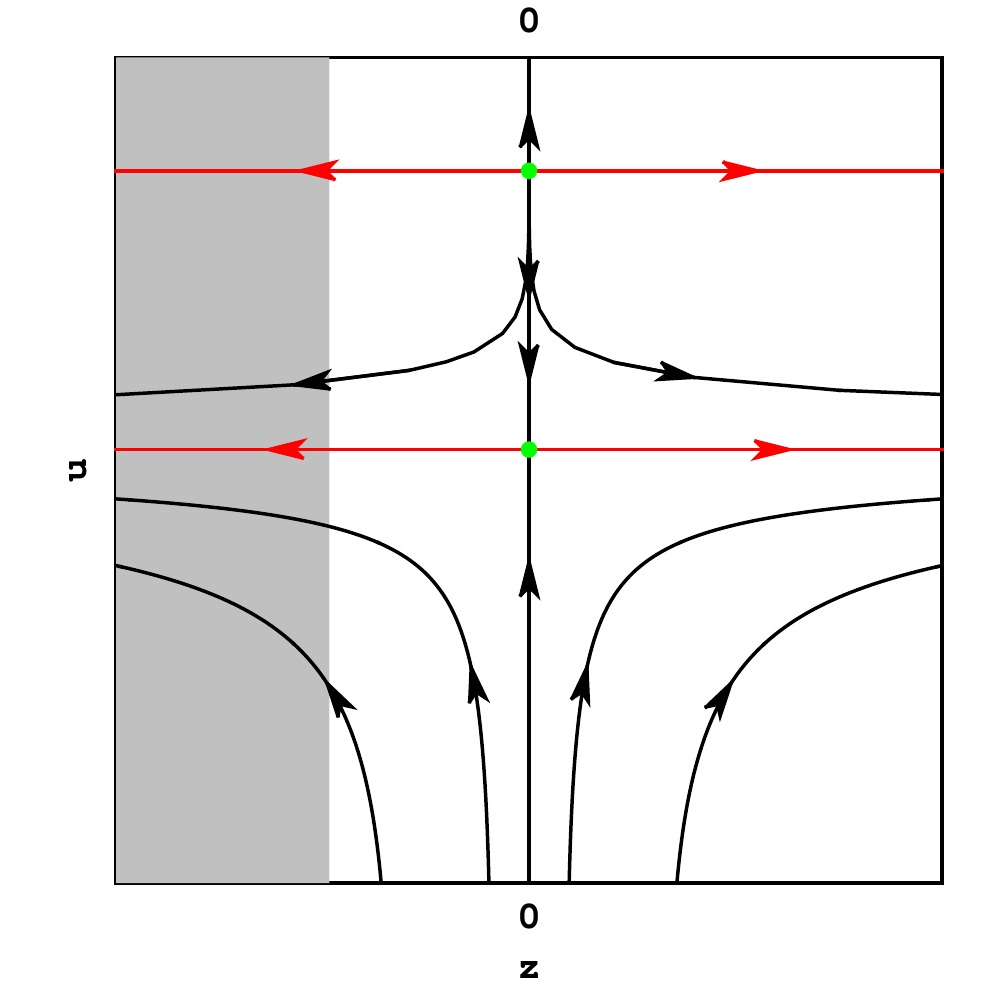}
\caption{The phase portrait of the system in the projective coordinates $(z,u)$ dedicated for the analysis of the behavior of trajectories at the infinity $H^2 \to \infty$. There are two critical points: a saddle point representing the Milne universe $H^2 \propto a^{-2}$ and an unstable node representing the Einstein-de Sitter-like universe. It is assumed the parameter $\varepsilon$ is rather small. The shadowed region is representing the non-physical domain forbidden for trajectories where $H^2 <0$.}
\label{fig:2}
\end{figure}

\begin{figure}
\centering
\includegraphics[width=0.8\textwidth]{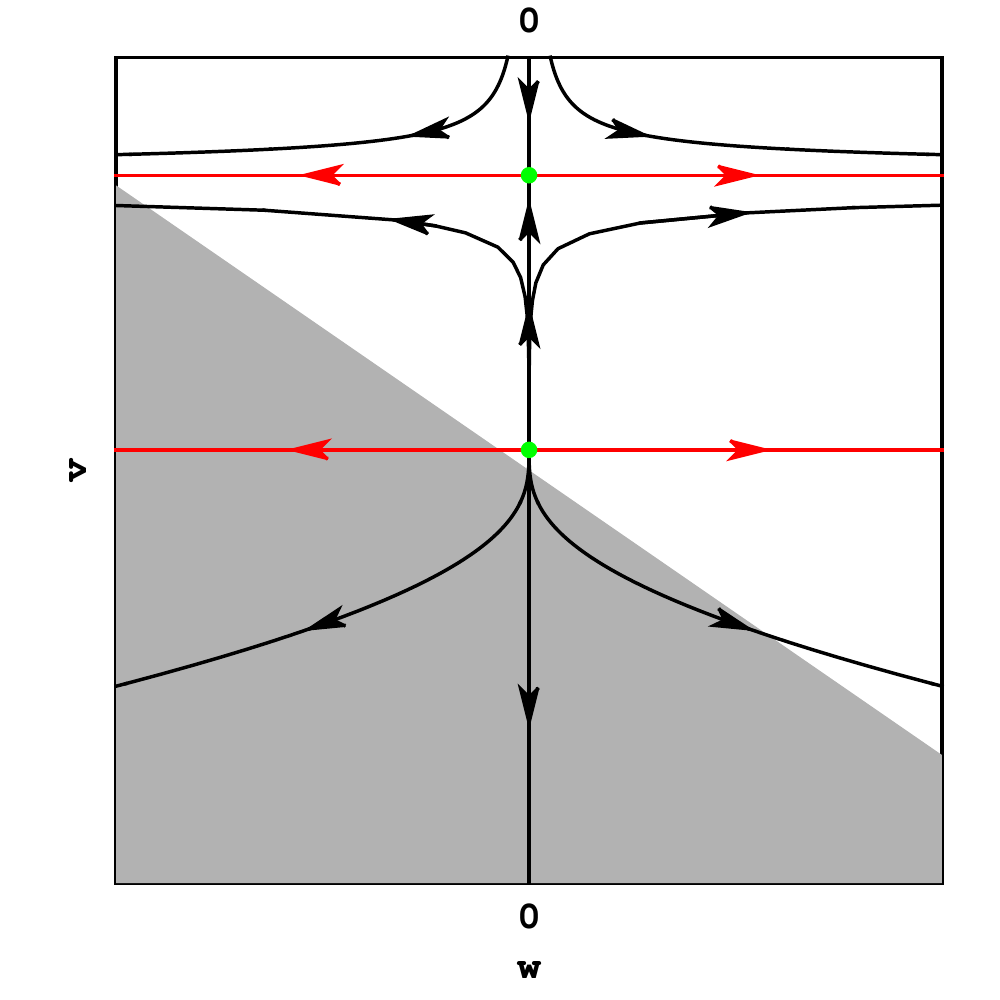}
\caption{The phase portrait of the system in the projective coordinates $(v,w)$ dedicated for the analysis of the behavior of trajectories at the infinity $H^2 \to \infty$. There are two critical points: a saddle point representing the Milne universe $H^2 \propto a^{-2}$ and an unstable node representing the Einstein-de Sitter-like universe. It is assumed the parameter $\varepsilon$ is rather small. The shadowed region is representing the non-physical domain forbidden for trajectories where $H^2 <0$.}
\label{fig:3}
\end{figure}

The phase portrait of the system under consideration are presented in Fig.~\ref{fig:2} and \ref{fig:3}. From these phase portraits one can observe how trajectories behave near the initial singularity. In both cases two critical points are situated on the lines $z=0$ and $v=0$. The phase portrait is structurally stable which means that the system is resistant with respect to small changes of its right-hand sides.

The above analysis can be continued for cosmological models with a general form $\Lambda=\Lambda(H)$. Recently Perico et al. have studied the dynamics and exact solution if $\Lambda(H)$ is given in power series with respect to $H$ \cite{Perico:2013mna}. The physical motivation for this studies can derived from quantum mechanics principles \cite{Urbanowski:2015ska}.

\section{Data}

To estimate the parameters of the both models we used the modified for our purposes \textsc{CosmoMC} code \cite{CosmoMC,Lewis:2002ah} with the implemented nested sampling algorithm \textsc{multinest} \cite{Feroz:2007kg,Feroz:2008xx}.

We used the observational data of 580 supernovae type Ia (the \textsc{Union2.1} compilation \cite{Suzuki:2011hu}), 31 observational data points of Hubble function \cite{Chen:2013vea}, the measurements of BAO (barion acoustic oscillations) from the Sloan Digital Sky Survey (SDSS-III) combined with the 2dF Galaxy Redshift Survey (2dFGRS) \cite{Eisenstein:2005su,Percival:2009xn,Eisenstein:2011sa,Ahn:2013gms}, the 6dF Galaxy Survey (6dFGS) \cite{Jones:2009yz,Beutler:2011hx}, the WiggleZ Dark Energy Survey \cite{Drinkwater:2009sd,Blake:2011en,Blake:2011wn}. We also used information coming from determinations of the Hubble function using the Alcock-Paczy\'{n}ski test \cite{Alcock:1979mp,Blake:2011ep}. This test is very restrictive in the context of modified gravity models.

We use the following likelihood functions for aforementioned data.

First, the likelihood function for the supernovae type Ia data
\begin{equation}
L_{\text{SN}} \propto \exp \left[ - \sum_{i,j}(\mu_{i}^{\text{obs}} - \mu_{i}^{\text{th}}) C_{ij}^{-1} (\mu_{j}^{\text{obs}} - \mu_{j}^{\text{th}})\right] , \label{sn_likelihood}
\end{equation}
where $C_{ij}$ is the covariance matrix with the systematic errors, $\mu_{i}^{\text{obs}}=m_{i}-M$ is the distance modulus, $\mu_{i}^{\text{th}}=5\log_{10}D_{Li} + \mathcal{M}=5\log_{10}d_{Li} + 25$, $\mathcal{M}=-5\log_{10}H_{0}+25$ and $D_{Li}=H_{0}d_{Li}$, where $d_{Li}$ is the luminosity distance which is given by $d_{Li}=(1+z_{i})c\int_{0}^{z_{i}} \frac{dz'}{H(z')}$ (with the assumption $k=0$).

Second, the likelihood function for $H(z)$ data
\begin{equation}
L_{H_z} \propto \exp \left[ - \sum_i\frac{\left(H^{\text{th}}(z_i)-H^{\text{obs}}_i\right)^2}{2 \sigma_i^2} \right ],
\label{hz_likelihood}
\end{equation}
where $H^{\text{th}}(z_i)$ denotes the theoretically estimated Hubble function, $H^{\text{obs}}_i$ is observational data.

Third, the likelihood function for the BAO data
\begin{equation}
L_{\text{BAO}} \propto \exp \left[ - \sum_{i,j}\left(d^{\text{th}}(z_i)-d^{\text{obs}}_i\right) C_{ij}^{-1} \left(d^{\text{th}}(z_j)-d^{\text{obs}}_j\right)\right] \label{bao_likelihood}
\end{equation}
where $C_{ij}$ is the covariance matrix with the systematic errors, $d^{\text{th}}(z_i)\equiv r_s(z_d) \left[(1+z_i)^2 D_A^2(z_i)\frac{cz_i}{H(z_i)} \right ]^{-\frac{1}{3}}$, $r_s(z_d)$ is the sound horizon at the drag epoch and $D_A$ is the angular diameter distance.

Fourth, the likelihood function for the information coming from the Alcock--Paczy\'{n}ski test
\begin{equation}
L_{AP} \propto \exp \left[ - \sum_i\frac{\left(AP^{\text{th}}(z_i)-AP^{\text{obs}}_i\right)^2}{2 \sigma_i^2} \right]
\label{ap_likelihood}
\end{equation}
where $AP^{\text{th}}(z_i)\equiv \frac{H(z_i)}{H_0 (1+z_i)}$.

Fifth, the likelihood function for the CMB shift parameter $R$ \cite{Bond:1997wr}
\begin{equation}
L_{\text{CMB}} \propto \exp \left[ -\frac{1}{2}\frac{(R^{\text{th}}-R^{\text{obs}})^2}{\sigma_{\mathcal{A}}^2} \right]
\label{cmbr_likelihood}
\end{equation}
where $R^{\text{th}}=\frac{\sqrt{\Omega_{\text{m}} H_0}}{c}(1+z_{*})D_\mathcal{A}(z_{*})$, $D_\mathcal{A} (z_{*})$ is the angular diameter distance to the last scattering surface, $R^{\text{obs}}=1.7477$ and $\sigma_{\mathcal{A}}^{-2}=48976.33$ \cite{Li:2013awa}.

And finally, the total likelihood function $L_{\text{tot}}$
\begin{equation}
L_{\text{tot}}=L_{\text{SN}}L_{H_z}L_{\text{BAO}}L_{\text{CMB}}L_{AP}.
\label{total_likelihood}
\end{equation}

\section{Statistical analysis}

\subsection{Estimation of model parameters}

Let us assume that we have $N$ pairs of measurements $(y_i,x_i)$ and that we want to find the relation between the $y$ and $x$ variables. Suppose that we can postulate $k$ possible relations $y\equiv f_i(x,\bar{\theta})$, where $\bar{\theta}$ is the vector of unknown model parameters and $i=1,\dots,k$. With the assumption that our observations come with uncorrelated Gaussian errors with a mean $\mu_i=0$ and a standard deviation $\sigma_i$ the goodness of fit for the theoretical model is measured by the $\chi^2$ quantity given by
\begin{equation}
\chi^2=\sum_{i=1}^{N} \frac{(f_l(x_i,\bar{\theta}) - y_i)^2}{2\sigma_i^2}=-2\ln L,
\end{equation}
where $L$ is the likelihood function. For the particular family of models $f_l$ the best one minimize the $\chi^2$ quantity, which we denote $f_l(x,\hat{\bar{\theta}})$. The best model from our set of $k$ models ${f_1(x,\hat{\bar{\theta}}),\dots,f_k(x,\hat{\bar{\theta}})}$ could be the one with the smallest value of $\chi^2$ quantity. But this method could give us misleading results. Generally speaking for more complex model the value of $\chi^2$ is smaller, thus the most complex one will be choose as the best from our set under consideration.

\begin{table*}
\scriptsize
\caption{Mean of marginalized posterior PDF with $68\%$ confidence level for the parameters of the models. In the brackets are shown parameter's values of joined posterior probabilities. Estimations were made using the Union2.1, $h(z)$, BAO, determinations of Hubble function using Alcock--Paczy\'{n}ski test and CMB R data sets.}
\begin{tabular*}{\textwidth}{@{\extracolsep{\fill}}lccccc@{}}
   \hline 
  & Union2.1 data & Union2.1, $h(z)$, BAO, A-P & Union2.1, $h(z)$, BAO, A-P, CMB \\
  \hline
  \multicolumn{4}{c}{testing model}\\
  \hline \hline 
    $\Omega_{\text{m},0} \in [0,1]$    & $0.3032^{+0.0164}_{-0.0165} (0.2916)$ & $0.2880^{+0.0160}_{-0.0174} (0.2814)$ & $0.2798^{+0.0096}_{-0.0104} (0.2818)$ \\
    $\Omega_{\Lambda,0} \in [0,1]$ & $0.6719^{+0.0712}_{-0.0753} (0.7069)$ & $0.6968^{+0.0386}_{-0.0395} (0.7059)$ & $0.7177^{+0.0092}_{-0.0082} (0.7159)$ \\
    $h_{100} \in [0.60,0.80]$ & $0.6998^{+0.1002}_{-0.0998} (0.7395)$ & $0.6917^{+0.0065}_{-0.0064} (0.6955)$ & $0.6927^{+0.0073}_{-0.0073} (0.6897)$ \\[2pt]
  \hline
  \multicolumn{4}{c}{$\Lambda$CDM model}\\
  \hline \hline
    $\Omega_{\text{m},0} \in [0,1]$    & $0.2953^{+0.0035}_{-0.0035} (0.2954)$ & $0.2777^{+0.0069}_{-0.0072} (0.2795)$ & $0.2908^{+0.0043}_{-0.0045} (0.2906)$ \\
    $h_{100} \in [0.60,0.80]$ & $0.7004^{+0.0996}_{-0.1004} (0.7789)$ & $0.6932^{+0.0050}_{-0.0050} (0.6917)$ & $0.6860^{+0.0042}_{-0.0041} (0.6868)$ \\[2pt]
  \hline
\end{tabular*}
\label{tab1_values}
\end{table*}

\subsection{The likelihood ratio test}

We perform the likelihood ratio test at the beginning of our statistical analysis. In this test one of the models (null model) is nested in a second model (alternative model) by fixing one of the second model parameters. In our case the null model is the $\Lambda$CDM model, the alternative model is the decaying vacuum $\Lambda$CDM model, and the parameter in question is $\Omega_{\text{vac}}$.
\begin{align*}
H_0 &\colon \Omega_{\text{m}} + \Omega_{\Lambda} = 1 \\
H_1 &\colon \Omega_{\text{m}} + \Omega_{\Lambda} \neq 1.
\end{align*}
The statistic is given by
\begin{equation}\label{lik_ratio_test}
\lambda = 2 \ln \left( \frac{L(H_1|D)}{L(H_0|D)} \right) = 2\left( \frac{\chi^2_{\text{vac}}}{2} - \frac{\chi^2_{\Lambda\text{CDM}}}{2} \right)
\end{equation}
where $L(H_1|D)$ is the likelihood of the decaying vacuum model, $L(H_0|D)$ is the likelihood of the $\Lambda$CDM model. The statistic $\lambda$ has the $\chi^2$ distribution with $df=n_1-n_0=2$ degree of freedom where $n_1$ is number of the parameters of the alternative model, $n_0$ is number of the parameters of the null model. The results are presented in Table~\ref{classic_test}. The three different sets of data have been used (the Union2.1 only, all data without CMB and all data). In all three cases the p-values are greater than the significance level $\alpha = 0.05$, that why the null hypothesis cannot be rejected. In other words we cannot reject the hypothesis that there is no decaying vacuum.

\begin{table*}
\caption{The results of the likelihood ratio test for the $\Lambda$CDM model (null model) and the decaying vacuum model (alternative model). The values of $\chi^2_{\text{test}}$, $\chi^2_{\Lambda\text{CDM}}$, test statistic $\lambda$ and corresponding p-values ($df=4-2=2$). Estimations were made using the Union2.1, $h(z)$, BAO, determinations of the Hubble function using Alcock--Paczy\'{n}ski test, and CMB R data sets.}
\label{classic_test}
\begin{tabular*}{\textwidth}{@{\extracolsep{\fill}}lcccc@{}}
 \hline
data sets &$\chi^2_{\text{vac}}/2$ & $\chi^2_{\Lambda\text{CDM}}/2$ & $\lambda$ & p-value \\
  \hline
Union2.1 & $272.5544$ & $272.5552$ & $0.0016$ & $0.9992$ \\
Union2.1, $h(z)$, BAO, AP & $282.2388$ & $282.2555$ & $0.0334$ & $0.9834$ \\
Union2.1, $h(z)$, BAO, AP, CMB & $282.2489$ & $282.4913$ & $0.4848$ & $0.7847$ \\[2pt]
  \hline
\end{tabular*}
\end{table*}

\subsection{Bayesian analysis of the models}

For the further comparison of the decaying vacuum model with the $\Lambda$CDM model we use the Bayesian statistical methods. 
These methods of models selection are widely used for cosmological model comparison \cite{Liddle:2004nh, Mukherjee:2005tr, Trotta:2005ar, Parkinson:2004yx, Liddle:2006kn, Parkinson:2006ku, Trotta:2007hy, Liddle:2007fy, Kurek:2007tb}. We should be aware that conclusions based on such analyses depend on the data at hand. New data obtained in future may change our conclusions.

First we calculate the Akaike information criterion (AIC) \cite{Akaike:1974nl} which is an approximation to the Kulback-Leibler information which measures information lost when an unknown (hypothetical) true model is approximated by a tested model. The AIC quantity is lowest for a best model and is given by
\begin{equation}
\text{AIC}=-2\ln \mathcal{L} +2d,
\end{equation}
where $\mathcal{L}$ is the maximum of the likelihood function and $d$ is the number of model parameters. It is convenient to evaluate the differences between the AIC quantities computed for the rest of models from the set and the AIC for the best one. Those differences ($\Delta_{\text{AIC}}$) are interpreted as a `strength of evidence' for a model considered with respect to the best one (i.e. with a lowest value of AIC). The models with $0 \le \Delta_{\text{AIC}}\le 2$ have substantial support (evidence), those where $4<\Delta_{\text{AIC}}\le 7$ have considerably less support, while models having $\Delta_{\text{AIC}} > 10 $ have essentially no support with respect to the best model.

The complexity of the model is interpreted here as the number of its free parameters that can be adjusted to fit the model to the observations. If models fit the data equally well according to the Akaike rule the best one is with the smallest number of model parameters (the simplest one in such an approach).

In the Bayesian framework the best model (from a model set under consideration) is that which has the largest value of probability in the light of data (so called posterior probability) \cite{Jeffreys:1961}
\begin{equation}
P(M_{i}|D)=\frac{P(D|M_{i})P(M_{i})}{P(D)},
\end{equation}
where $P(M_{i})$ is a prior probability for the model $M_{i}$ (it is assumed that there is no evidence to favor one model over another, so values of priors for all models under consideration are equal), $D$ denotes data, $P(D)$ is the normalization constant
\begin{equation}
P(D)= \sum _{i=1}^{k} P(D|M_{i})P(M_{i}).
\end{equation}
And $P(D|M_{i})$ is the marginal likelihood, also called evidence
\begin{equation}
P(D|M_{i})=\int P(D|\bar{\theta},M_{i})P(\bar{\theta}|M_{i}) \ d \bar{\theta} \equiv E_{i},
\end{equation}
where $P(D|\bar{\theta},M_{i})$ is likelihood under model $i$, $P(\bar{\theta}|M_{i})$ is prior probability for ${\bar{\theta}}$ under model $i$.

It is convenient to evaluate the posterior ratio for models under consideration which in the case with flat prior for models is reduced to the evidence ratio called the Bayes factor
\begin{equation}
B_{ij} = \frac{P(D|M_i)}{P(D|M_j)}.
\end{equation}
The interpretation of twice the natural logarithm of the Bayes factor is as follow: $0<2\ln B_{ij}\leq 2$ as a weak evidence, $2<2\ln B_{ij}\leq 6$ as a positive evidence, $6<2\ln B_{ij}\leq 10$ as a strong evidence and $2\ln B_{ij}> 10$ as a very strong evidence against model $j$ comparing to model $i$. Let us simplify the problem to illustrate how this principle works here \cite{MacKay:2003it,Trotta:2005ar}.

Assume that $\bar{P}(\bar{\theta}|D,M)$ is the non normalized posterior probability for the vector $\bar{\theta}$ of model parameters. In this notation $E=\int\bar{P}(\bar{\theta}|D,M)d\bar{\theta}$. Suppose that posterior has a strong peak in the maximum: $\bar{\theta}_{\text{MOD}}$. It is reasonable to approximate the logarithm of the posterior by its Taylor expansion in the neighborhood of $\bar{\theta}_{\text{MOD}}$ so we finished with the expression
\begin{align}
\bar{P}(\bar{\theta}|D,M) & = \bar{P}(\bar{\theta}_\text{MOD}|D,M) \times \nonumber \\
& \times \exp \left[-(\bar{\theta}-\bar{\theta}_{\text{MOD}})^T C^{-1}(\bar{\theta}-\bar{\theta}_{\text{MOD}})\right],
\end{align}
where $\left[ C^{-1} \right]_{ij} = -\left[\frac{\partial^2\ln\bar{P}(\bar{\theta}|D,M)}{\partial\theta_i\partial\theta_j}\right]_{\bar{\theta}=\bar{\theta}_\text{MOD}}$. The posterior is approximated by the Gaussian distribution with the mean $\bar{\theta}_\text{MOD}$ and the covariance matrix $C$. The evidence then has a form
\begin{eqnarray}
E & = & \bar{P}(\bar{\theta}_{\text{MOD}}|D,M) \times \nonumber \\
& \times & \int \exp \left[-(\bar{\theta}-\bar{\theta}_{\text{MOD}})^T C^{-1}(\bar{\theta}-\bar{\theta}_{\text{MOD}})\right] \ d \bar{\theta}.
\end{eqnarray}
Because the posterior has a strong peak near the maximum, the most contribution to the integral comes from the neighborhood close to $\bar{\theta}_\text{MOD}$. Contribution from the other region of $\bar{\theta}$ can be ignored, so we can expand the limit of the integral to whole $R^d$. With this assumption one can obtain $E=(2\pi)^{\frac{d}{2}}\sqrt{\det C}\bar{P}(\bar{\theta}_{\text{MOD}}|D,M)= (2\pi)^{\frac{d}{2}}\sqrt{\det C}P(D|\bar{\theta}_\text{MOD},M)P(\bar{\theta}_\text{MOD}|M)$.
Suppose that the likelihood function has a sharp peak in $\hat{\bar{\theta}}$ and the prior for $\bar{\theta}$ is nearly flat in the neighborhood of $\hat{\bar{\theta}}$. In this case $\hat{\bar{\theta}}=\bar{\theta}_{\text{MOD}}$ and the expression for the evidence takes the form $E=\mathcal{L}(2\pi)^{\frac{d}{2}}\sqrt{\det\text{C}}P(\hat{\bar{\theta}}|M)$. The quantity $(2\pi)^{\frac{d}{2}}\sqrt{\det\text{C}}P(\hat{\bar{\theta}}|M)$ is called the Occam factor (OF). When we consider the case with one model parameter with a flat prior $P(\theta|M)=\frac{1}{\Delta\theta}$ the Occam factor OF$=\frac{2\pi\sigma}{\Delta\theta}$ which can be interpreted as the ratio of the volume occupied by the posterior to the volume occupied by prior in the parameter space. The more parameter space wasted by the prior the smaller value of the evidence. It is worth noting that the evidence does not penalize parameters which are unconstrained by the data \cite{Liddle:2006kn}.

As the evidence is hard to evaluate an approximation to this quantity was proposed by Schwarz \cite{Schwarz:1978ed} so called Bayesian information criterion (BIC) and is given by
\begin{equation}
\text{BIC}=-2\ln\mathcal{L}+2d\ln N,
\end{equation}
where $N$ is the number of the data points. The best model from a set under consideration is this which minimizes the BIC quantity. One can notice the similarity between the AIC and BIC quantities though they come from different approaches to model selection problem. The dissimilarity is seen in the so called penalty term: $ad$, which penalize more complex models (complexity is identified here as the number of free model parameters). One can evaluated the factor by which the additional parameter must improve the goodness of fit to be included in the model. This factor must be greater than $a$ so equal to $2$ in the AIC case and equal to $\ln N$ in the BIC case. Notice that the latter depends on the number of the data points.

It can be shown that there is the simple relation between the BIC and the Bayes factor
\begin{equation}
    2 \ln B_{ij} = -(\text{BIC}_i - \text{BIC}_j).
\end{equation}
The quantity $B_{ij}$ is the Bayes factor for the hypothesis (model) $i$ against the hypothesis (model) $j$. We categorize this evidence against the model $j$ taking the following ranking. The evidence against the model $j$ is not worth than bare mention when twice the natural logarithm of the Bayes factor (or minus the difference between BICs) is $0< 2\ln B_{ij} \leq 2$, is positive when $2< 2\ln B_{ij} \leq 6$, is strong when $6< 2\ln B_{ij}\leq 10$ and is very strong when $ 2\ln B_{ij} > 10$.

To obtain the values of AIC and BIC quantities we perform the $\chi^2=-2\ln L$ minimization procedure after marginalization over the $H_0$ parameter in the range $[60, 80]$. They are presented in Table~\ref{aic_bic}.

\begin{table*}
\scriptsize
\caption{Values of $\chi^2$, AIC, BIC and Bayes factor, $\Delta$AIC (with respect to the $\Lambda$CDM model). Estimations were made using the Union2.1, $h(z)$, BAO, determinations of Hubble function using Alcock--Paczy\'{n}ski test and CMB R data sets.}
\begin{tabular*}{\textwidth}{@{\extracolsep{\fill}}lccccccc@{}}
  \hline
  &$\chi^2/2$ & Evidence $\ln E_i$ & AIC & $\text{AIC}_{j}-\text{AIC}_{i}$ & BIC & $2\ln B_{ij}$ \\
  \hline
  \multicolumn{7}{c}{tested model}\\
  \hline
Union2.1 & $272.5544$ & $-275.1931^{+0.0847}_{-0.1248}$ & $551.1088$ & $1.9984$ & $564.1979$ & $6.3614$ \\
Union2.1, $h(z)$, BAO, A-P & $282.2388$ & $-289.0193^{+0.0900}_{-0.1387}$ & $570.4776$ & $1.9666$ & $583.7716$ & $6.3979$ \\
Union2.1, $h(z)$, BAO, A-P, CMB & $282.2489$ & $-292.4944^{+0.0726}_{-0.1452}$ & $570.4978$ & $1.5152$ & $583.7966$ & $5.9481$ \\
  \hline
  \multicolumn{7}{c}{$\Lambda$CDM model}\\
  \hline
Union2.1 & $272.5552$ & $-274.7896^{+0.1110}_{-0.0620}$ & $549.1104$ & --- & $557.8365$ & --- \\
Union2.1, $h(z)$, BAO, A-P & $282.2555$ & $-287.1432^{+0.1152}_{-0.0942}$ & $568.5110$ & --- & $577.3736$ & --- \\
Union2.1, $h(z)$, BAO, A-P, CMB & $282.4913$ & $-287.9753^{+0.1415}_{-0.2909}$ & $568.9826$ & --- & $577.8485$ & --- \\
  \hline
\end{tabular*}
\label{aic_bic}
\end{table*}

\begin{figure}[h!!!!]
\centering
   \includegraphics[width=0.496\textwidth]{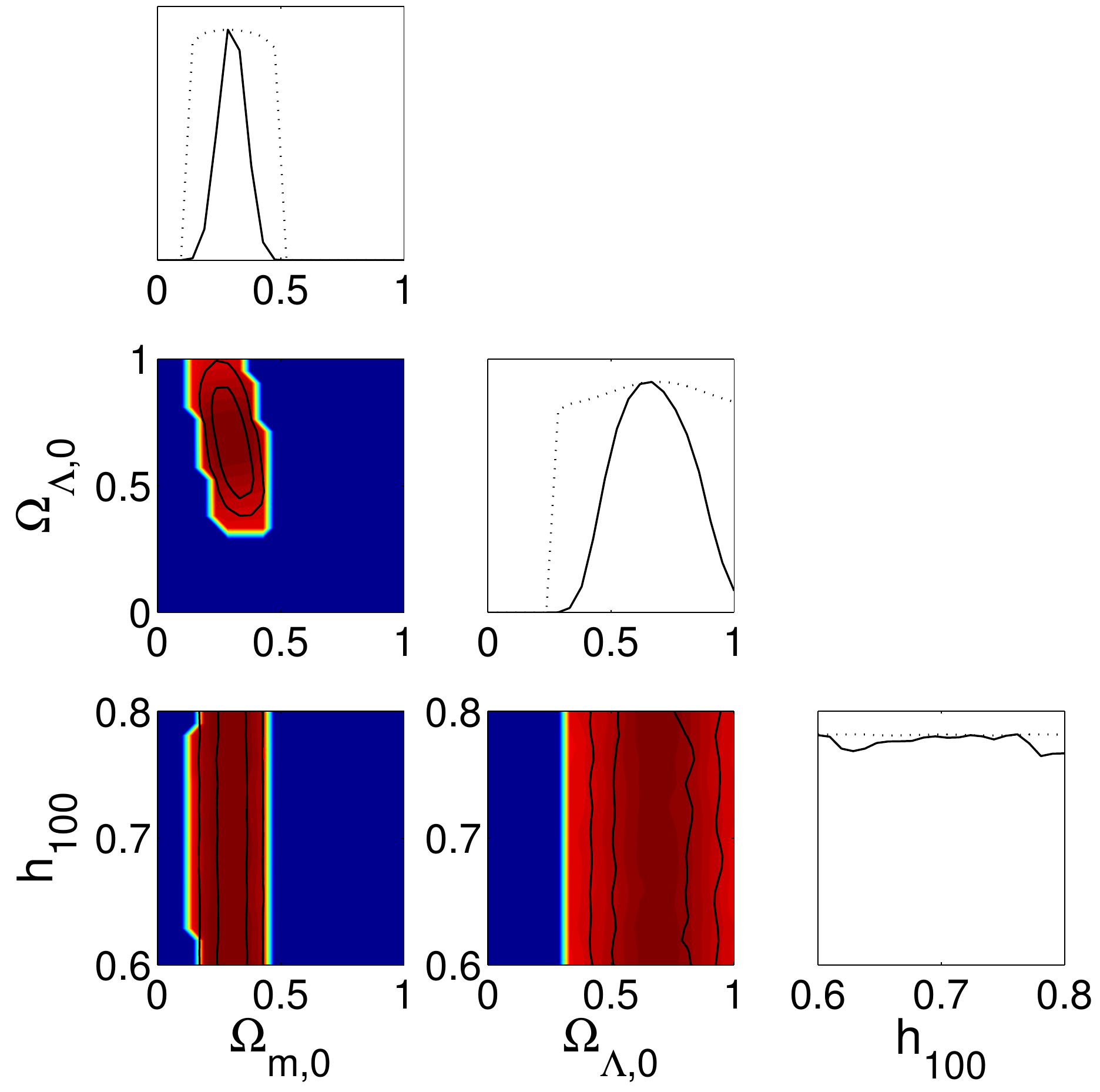}
   \includegraphics[width=0.496\textwidth]{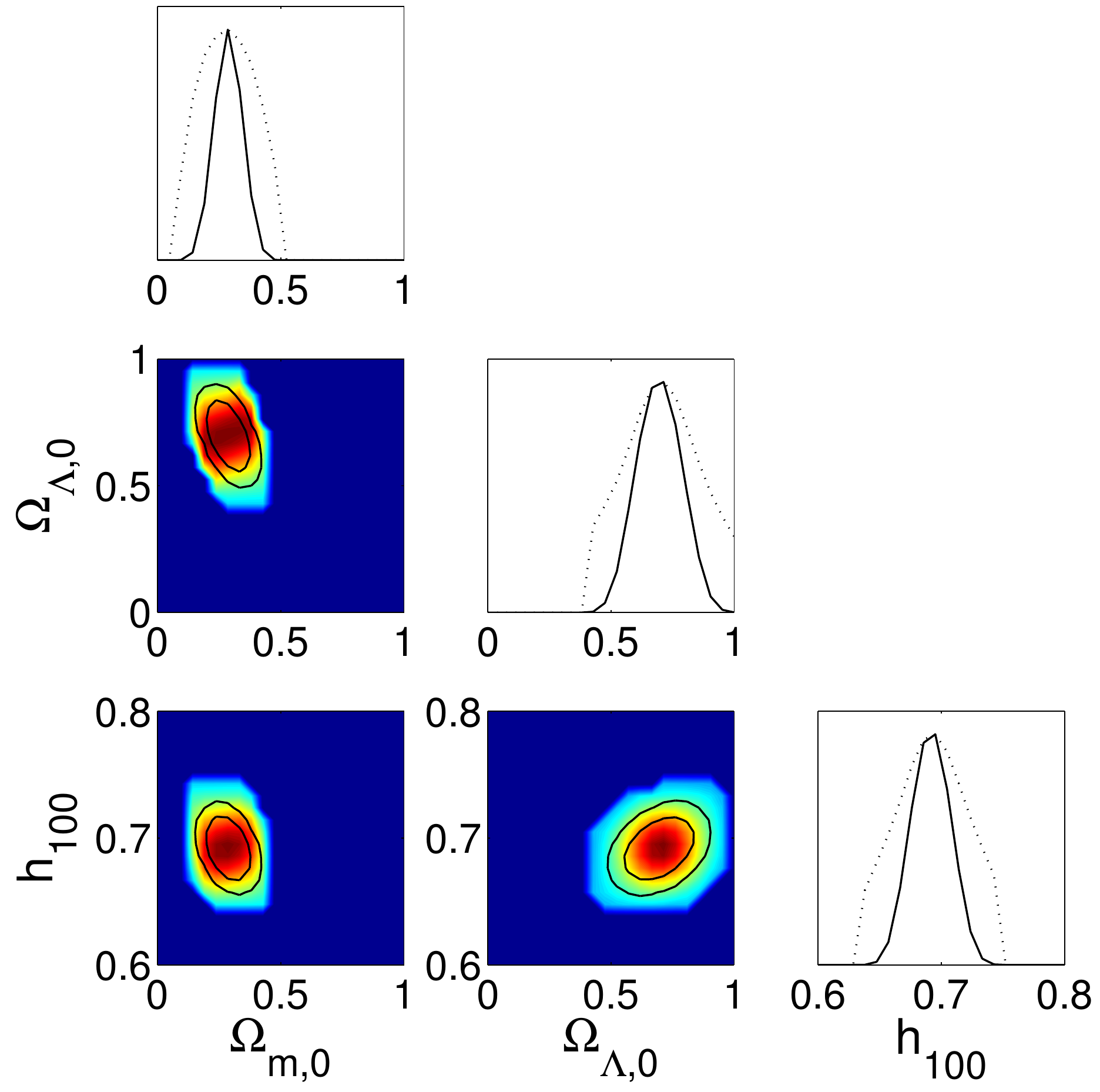}\\
   \includegraphics[width=0.496\textwidth]{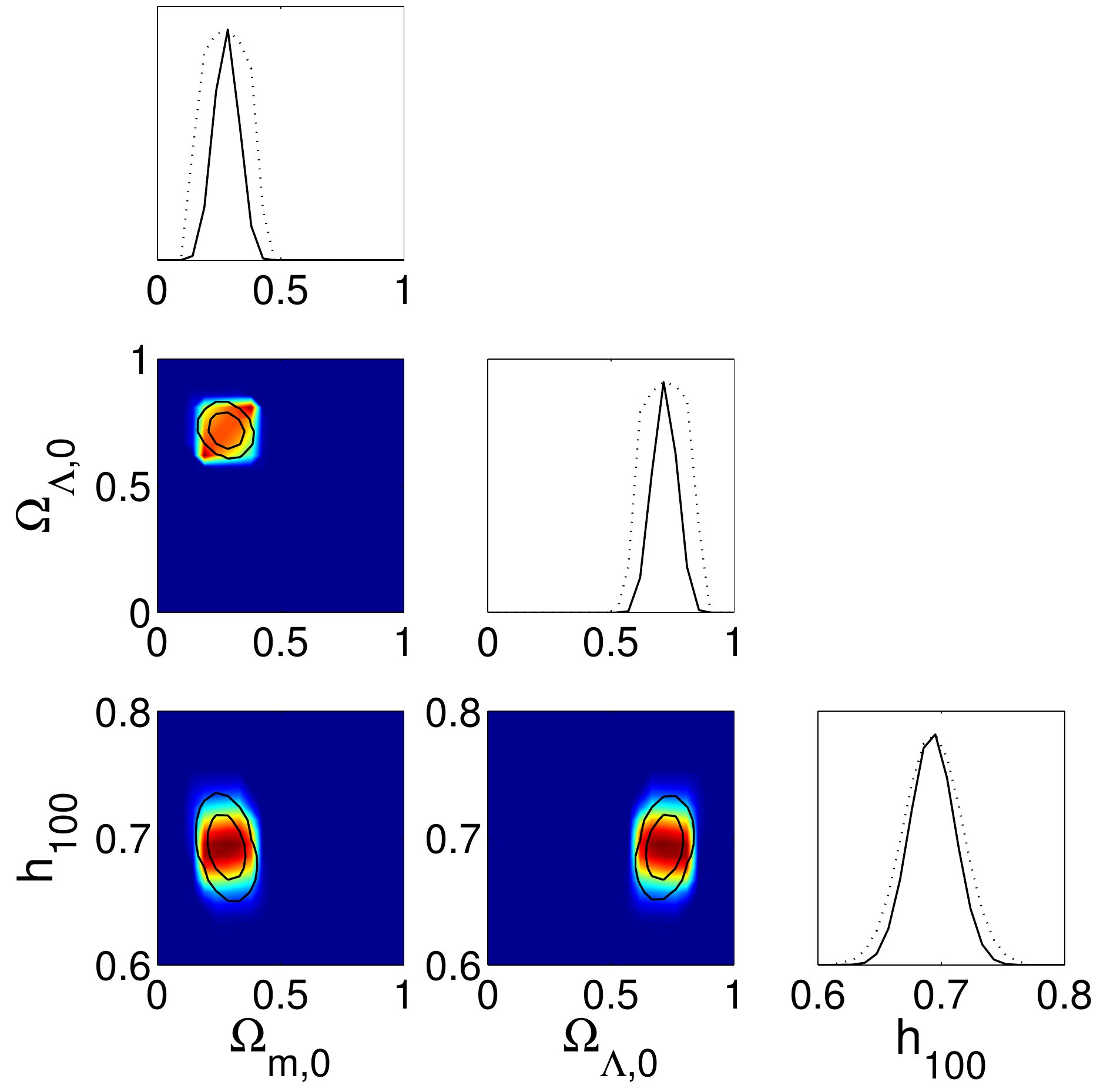}\\
  \caption{Posterior constraints for testing model. Solid lines denote $68\%$ and $95\%$ credible intervals of fully marginalized probabilities, the colors illustrate mean likelihood of the sample. Top left: estimations with the Union2.1 data only. Top right: estimations made using the Union2.1, $h(z)$, BAO, determinations of Hubble function using Alcock--Paczy\'{n}ski test data sets. Bottom: estimations made using the Union2.1, $h(z)$, BAO, determinations of Hubble function using Alcock--Paczy\'{n}ski test and CMB R data sets.}
   \label{pos_mod_int_de}
\end{figure}

\begin{figure}[h!!!!]
\centering
   \includegraphics[width=0.3\textwidth]{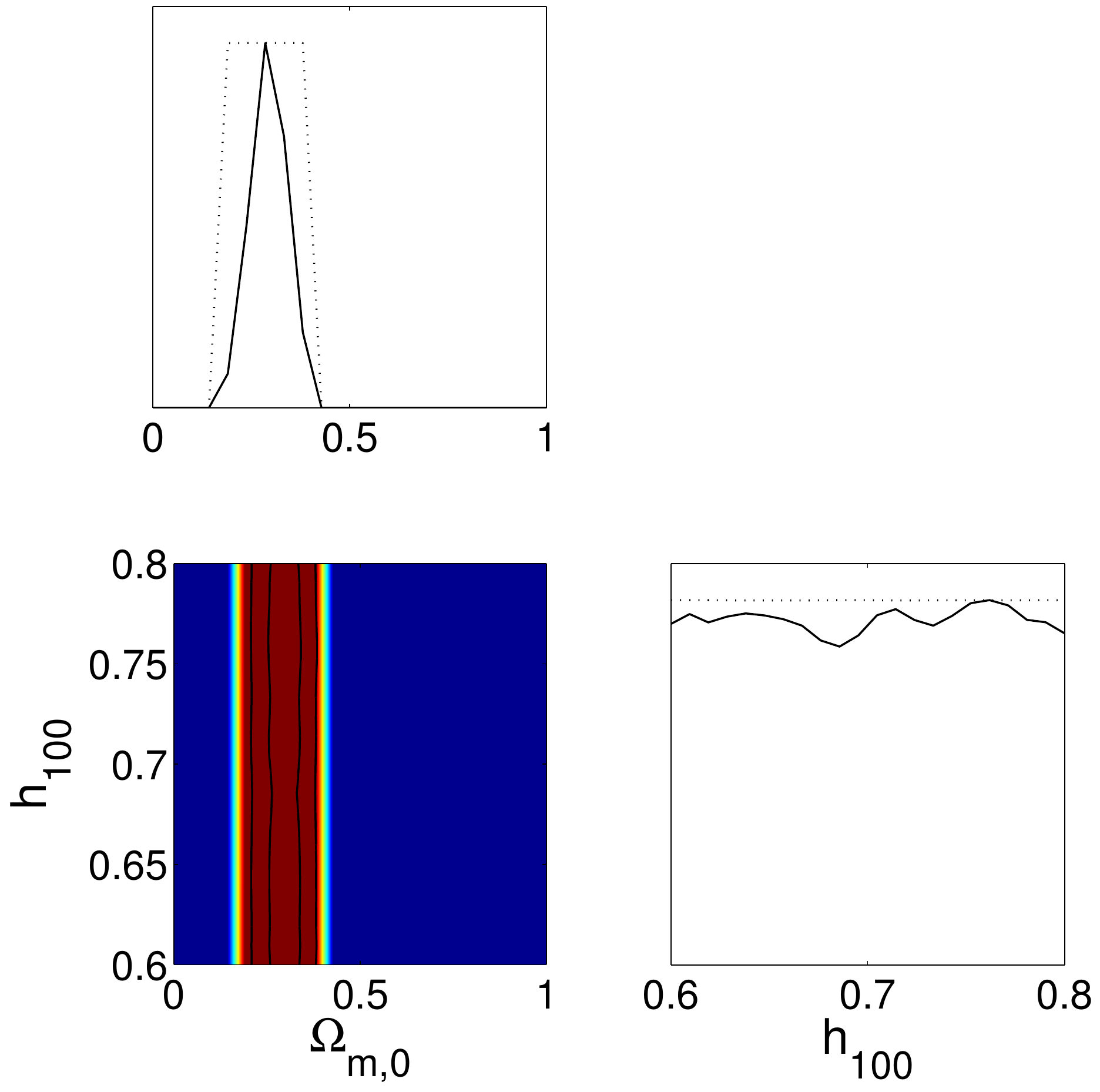}
   \includegraphics[width=0.3\textwidth]{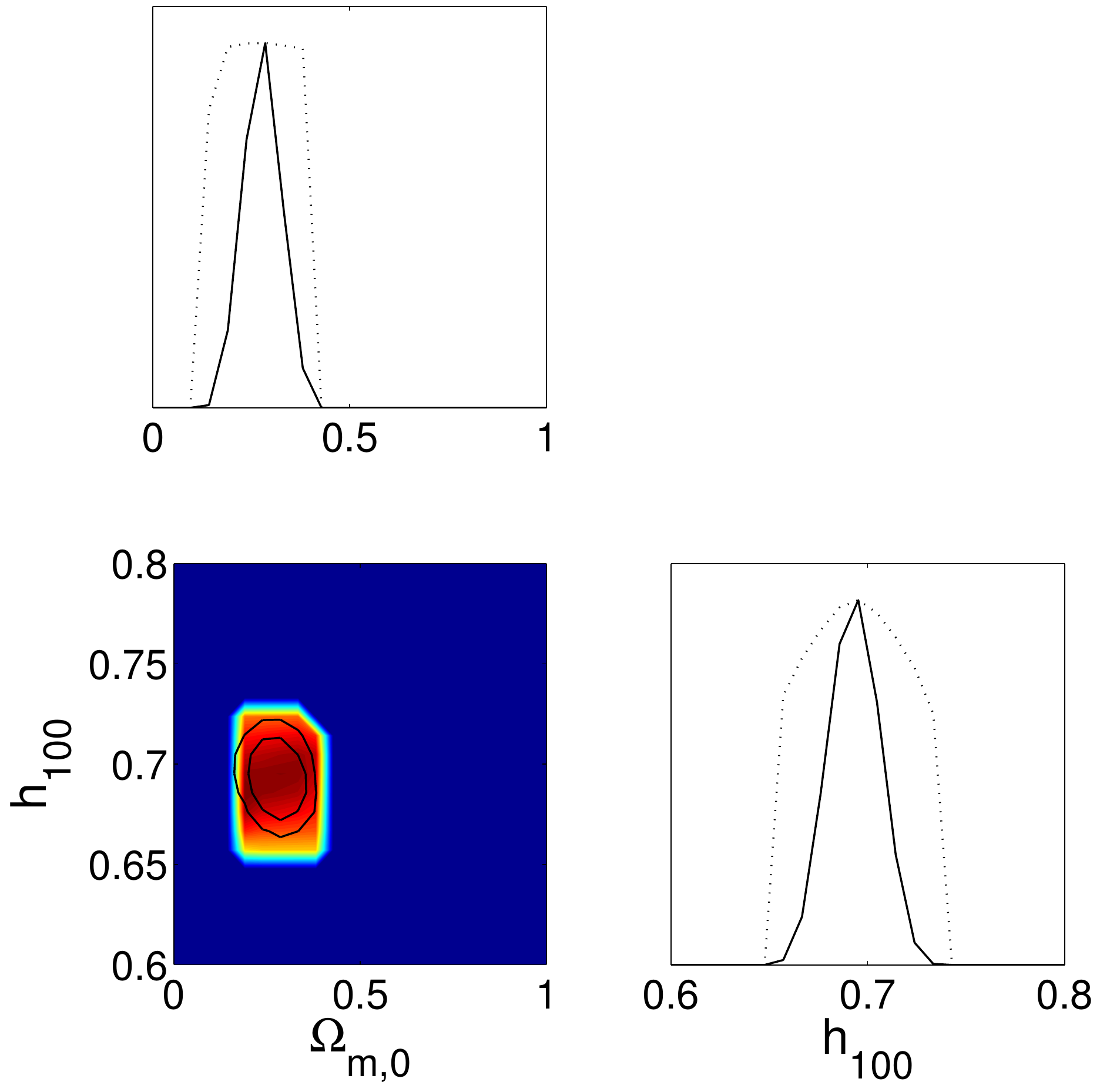}\\
   \includegraphics[width=0.3\textwidth]{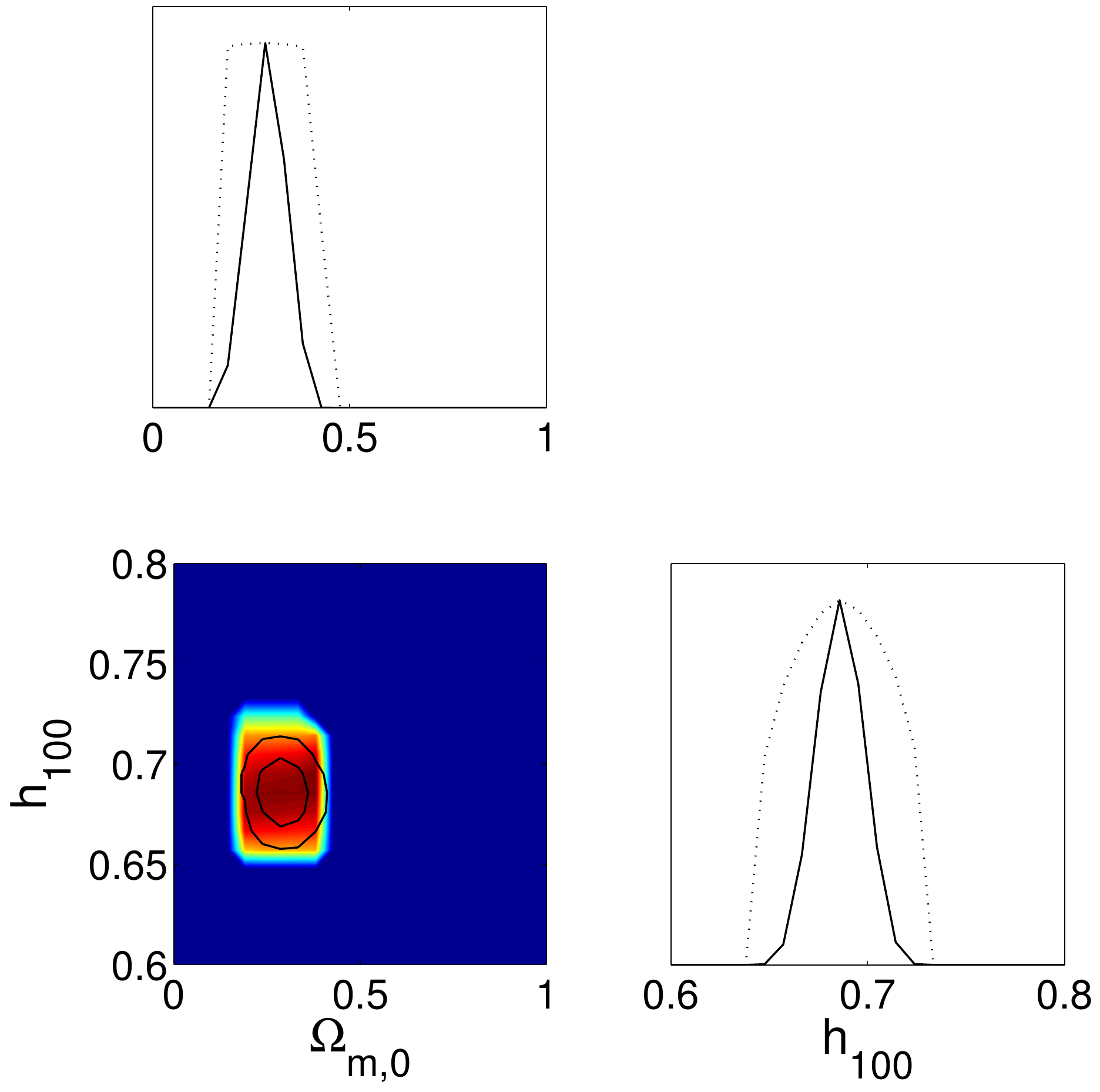}\\
   \caption{Posterior constraints for $\Lambda$CDM model. Solid lines denote $68\%$ and $95\%$ credible intervals of fully marginalized probabilities, the colors illustrate mean likelihood of the sample. Top left: estimations with the Union2.1 data only. Top right: estimations made using the Union2.1, $h(z)$, BAO, determinations of Hubble function using Alcock--Paczy\'{n}ski test data sets. Bottom: estimations made using the Union2.1, $h(z)$, BAO, determinations of Hubble function using Alcock--Paczy\'{n}ski test and CMB R data sets.}
   \label{pos_mod_lcdm}
\end{figure}

\section{Conclusion}

The main goal of the paper was to find the theoretical explanation of the cosmological constant problem. To solve the disrepancy of the value of the cosmological constant interpreted as vacuum energy and value of cosmological constant required for the explanation of acceleration of the current Universe we construct the cosmological model with the decaying vacuum. The parametrization of running cosmological constant is taken directly from quantum mechanics principles which distinguish the power-law relation for decaying part of vacuum dark energy.

Let us compare both the value of vacuum dark energy at the Planck epoch and today. Then we obtain
\begin{equation}
\frac{\rho_{\Lambda}(t \simeq t_{\text{Pl}})}{\rho_{\Lambda}(t\simeq t_0)} \simeq \frac{\alpha t_{\text{Pl}}^{-2}}{3H_{0}^{2}}
\end{equation} 
Expressing $H_0$ in the Planck unit ($H_0 = 1.24 \cdot 10^{-61} t_{\text{Pl}}^{-1}$) we obtain
\begin{equation}
\frac{\rho_{\Lambda}(t \simeq t_{\text{Pl}})}{\rho_{\Lambda}(t\simeq t_0)} \simeq \alpha \cdot 4.5 \cdot 10^{122}.
\end{equation} 
From the best fit value $1-\Omega_{\text{m},0}-\Omega_{\Lambda}=0.0025$ we found that 
\begin{equation}
\alpha = 3 (1-\Omega_{\text{m},0}-\Omega_{\Lambda}) = 0.0074 \simeq 10^{-2}
\end{equation}
and finally
\begin{equation}
\frac{\rho_{\Lambda}(t \simeq t_{\text{Pl}})}{\rho_{\Lambda}(t\simeq t_0)} \simeq 10^{120}.
\end{equation}

On the ground of our statistical analysis we can conclude that the cosmological model with decaying vacuum should be rejected and the $\Lambda$CDM model can keep a crown of the standard cosmological model. We showed that the cosmological model with decaying vacuum is worst-fitted than the $\Lambda$CDM model, but we think that it should not be considered yet as a blind alley. The strength of this model is its explanation of the cosmological constant conundrum. That why it should be considered seriously as a best alternative for the $\Lambda$CDM model from the physical point of view. 

We share George Efstathiou's opinion \cite{Efstathiou:2007gz,Chongchitnan:2007eb,Szydlowski:2004jv} that without sound theoretical basis there is no reason for considering the dynamical dark energy. This paper is the attempt to base the cosmological model with varying $\Lambda$ on the quantum mechanics principles. We submit this model for consideration although it is worse than standard cosmological model in the statistical terms. But it solves the cosmological constant problem. The $\Lambda$CDM model has only the status of the satisfactory effective theory. Our model has a deeper explanatory value and we think it indicates the direction of the future research in cosmology.

\acknowledgments
The work was supported by the grant NCN DEC-2013/09/B/ST2/03455. The author thanks K. Urbanowski, A. Krawiec, Z. Haba, M. Kamionka and A. Borowiec for interesting remarks and comments.


\begin{thebibliography}{52}
\expandafter\ifx\csname natexlab\endcsname\relax\def\natexlab#1{#1}\fi
\expandafter\ifx\csname bibnamefont\endcsname\relax
  \def\bibnamefont#1{#1}\fi
\expandafter\ifx\csname bibfnamefont\endcsname\relax
  \def\bibfnamefont#1{#1}\fi
\expandafter\ifx\csname citenamefont\endcsname\relax
  \def\citenamefont#1{#1}\fi
\expandafter\ifx\csname url\endcsname\relax
  \def\url#1{\texttt{#1}}\fi
\expandafter\ifx\csname urlprefix\endcsname\relax\def\urlprefix{URL }\fi
\providecommand{\bibinfo}[2]{#2}
\providecommand{\eprint}[2][]{\url{#2}}

\bibitem[{\citenamefont{Weinberg}(1989)}]{Weinberg:1988cp}
\bibinfo{author}{\bibfnamefont{S.}~\bibnamefont{Weinberg}},
  \bibinfo{journal}{Rev.Mod.Phys.} \textbf{\bibinfo{volume}{61}},
  \bibinfo{pages}{1} (\bibinfo{year}{1989}).

\bibitem[{\citenamefont{Alam et~al.}(2003)\citenamefont{Alam, Sahni, and
  Starobinsky}}]{Alam:2003rw}
\bibinfo{author}{\bibfnamefont{U.}~\bibnamefont{Alam}},
  \bibinfo{author}{\bibfnamefont{V.}~\bibnamefont{Sahni}}, \bibnamefont{and}
  \bibinfo{author}{\bibfnamefont{A.}~\bibnamefont{Starobinsky}},
  \bibinfo{journal}{JCAP} \textbf{\bibinfo{volume}{04}}, \bibinfo{pages}{002}
  (\bibinfo{year}{2003}), \eprint{astro-ph/0302302}.

\bibitem[{\citenamefont{Alcaniz and Lima}(2005)}]{Alcaniz:2005dg}
\bibinfo{author}{\bibfnamefont{J.~S.} \bibnamefont{Alcaniz}} \bibnamefont{and}
  \bibinfo{author}{\bibfnamefont{J.}~\bibnamefont{Lima}},
  \bibinfo{journal}{Phys.Rev.} \textbf{\bibinfo{volume}{D72}},
  \bibinfo{pages}{063516} (\bibinfo{year}{2005}), \eprint{astro-ph/0507372}.

\bibitem[{\citenamefont{Wang and Meng}(2005)}]{Wang:2004cp}
\bibinfo{author}{\bibfnamefont{P.}~\bibnamefont{Wang}} \bibnamefont{and}
  \bibinfo{author}{\bibfnamefont{X.-H.} \bibnamefont{Meng}},
  \bibinfo{journal}{Class.Quant.Grav.} \textbf{\bibinfo{volume}{22}},
  \bibinfo{pages}{283} (\bibinfo{year}{2005}), \eprint{astro-ph/0408495}.

\bibitem[{\citenamefont{Graef et~al.}(2014)\citenamefont{Graef, Costa, and
  Lima}}]{Graef:2013iia}
\bibinfo{author}{\bibfnamefont{L.}~\bibnamefont{Graef}},
  \bibinfo{author}{\bibfnamefont{F.}~\bibnamefont{Costa}}, \bibnamefont{and}
  \bibinfo{author}{\bibfnamefont{J.}~\bibnamefont{Lima}},
  \bibinfo{journal}{Phys.Lett.} \textbf{\bibinfo{volume}{B728}},
  \bibinfo{pages}{400} (\bibinfo{year}{2014}), \eprint{1303.2075}.

\bibitem[{\citenamefont{Coleman}(1977)}]{Coleman:1977py}
\bibinfo{author}{\bibfnamefont{S.~R.} \bibnamefont{Coleman}},
  \bibinfo{journal}{Phys.Rev.} \textbf{\bibinfo{volume}{D15}},
  \bibinfo{pages}{2929} (\bibinfo{year}{1977}).

\bibitem[{\citenamefont{Callan and Coleman}(1977)}]{Callan:1977pt}
\bibinfo{author}{\bibfnamefont{J.}~\bibnamefont{Callan},
  \bibfnamefont{Curtis~G.}} \bibnamefont{and}
  \bibinfo{author}{\bibfnamefont{S.~R.} \bibnamefont{Coleman}},
  \bibinfo{journal}{Phys.Rev.} \textbf{\bibinfo{volume}{D16}},
  \bibinfo{pages}{1762} (\bibinfo{year}{1977}).

\bibitem[{\citenamefont{Krauss and Dent}(2008)}]{Krauss:2007rx}
\bibinfo{author}{\bibfnamefont{L.~M.} \bibnamefont{Krauss}} \bibnamefont{and}
  \bibinfo{author}{\bibfnamefont{J.}~\bibnamefont{Dent}},
  \bibinfo{journal}{Phys.Rev.Lett.} \textbf{\bibinfo{volume}{100}},
  \bibinfo{pages}{171301} (\bibinfo{year}{2008}), \eprint{0711.1821}.

\bibitem[{\citenamefont{Urbanowski and Szydlowski}(2012)}]{Urbanowski:2012pka}
\bibinfo{author}{\bibfnamefont{K.}~\bibnamefont{Urbanowski}} \bibnamefont{and}
  \bibinfo{author}{\bibfnamefont{M.}~\bibnamefont{Szydlowski}},
  \bibinfo{journal}{AIP Conf.Proc.} \textbf{\bibinfo{volume}{1514}},
  \bibinfo{pages}{143} (\bibinfo{year}{2012}), \eprint{1304.2796}.

\bibitem[{\citenamefont{Urbanowski and Raczynska}(2014)}]{Urbanowski:2013tfa}
\bibinfo{author}{\bibfnamefont{K.}~\bibnamefont{Urbanowski}} \bibnamefont{and}
  \bibinfo{author}{\bibfnamefont{K.}~\bibnamefont{Raczynska}},
  \bibinfo{journal}{Phys.Lett.} \textbf{\bibinfo{volume}{B731}},
  \bibinfo{pages}{236} (\bibinfo{year}{2014}), \eprint{1303.6975}.

\bibitem[{\citenamefont{Lopez and Nanopoulos}(1996)}]{Lopez:1995eb}
\bibinfo{author}{\bibfnamefont{J.~L.} \bibnamefont{Lopez}} \bibnamefont{and}
  \bibinfo{author}{\bibfnamefont{D.~V.} \bibnamefont{Nanopoulos}},
  \bibinfo{journal}{Mod.Phys.Lett.} \textbf{\bibinfo{volume}{A11}},
  \bibinfo{pages}{1} (\bibinfo{year}{1996}), \eprint{hep-ph/9501293}.

\bibitem[{\citenamefont{Lima et~al.}(2015)\citenamefont{Lima, Perico, and
  Zilioti}}]{Lima:2015kda}
\bibinfo{author}{\bibfnamefont{J.}~\bibnamefont{Lima}},
  \bibinfo{author}{\bibfnamefont{E.}~\bibnamefont{Perico}}, \bibnamefont{and}
  \bibinfo{author}{\bibfnamefont{G.}~\bibnamefont{Zilioti}}
  (\bibinfo{year}{2015}), \eprint{1502.01913}.

\bibitem[{\citenamefont{Bessada and Miranda}(2013)}]{Bessada:2013maa}
\bibinfo{author}{\bibfnamefont{D.}~\bibnamefont{Bessada}} \bibnamefont{and}
  \bibinfo{author}{\bibfnamefont{O.~D.} \bibnamefont{Miranda}},
  \bibinfo{journal}{Phys.Rev.} \textbf{\bibinfo{volume}{D88}},
  \bibinfo{pages}{083530} (\bibinfo{year}{2013}), \eprint{1310.8571}.

\bibitem[{\citenamefont{Lima}(1996)}]{Lima:1995kd}
\bibinfo{author}{\bibfnamefont{J.}~\bibnamefont{Lima}},
  \bibinfo{journal}{Phys.Rev.} \textbf{\bibinfo{volume}{D54}},
  \bibinfo{pages}{2571} (\bibinfo{year}{1996}), \eprint{gr-qc/9605055}.

\bibitem[{\citenamefont{Dunajski and Gibbons}(2008)}]{Dunajski:2008tg}
\bibinfo{author}{\bibfnamefont{M.}~\bibnamefont{Dunajski}} \bibnamefont{and}
  \bibinfo{author}{\bibfnamefont{G.}~\bibnamefont{Gibbons}},
  \bibinfo{journal}{Class.Quant.Grav.} \textbf{\bibinfo{volume}{25}},
  \bibinfo{pages}{235012} (\bibinfo{year}{2008}), \eprint{0807.0207}.

\bibitem[{\citenamefont{Perico et~al.}(2013)\citenamefont{Perico, Lima,
  Basilakos, and Sola}}]{Perico:2013mna}
\bibinfo{author}{\bibfnamefont{E.}~\bibnamefont{Perico}},
  \bibinfo{author}{\bibfnamefont{J.}~\bibnamefont{Lima}},
  \bibinfo{author}{\bibfnamefont{S.}~\bibnamefont{Basilakos}},
  \bibnamefont{and} \bibinfo{author}{\bibfnamefont{J.}~\bibnamefont{Sola}},
  \bibinfo{journal}{Phys.Rev.} \textbf{\bibinfo{volume}{D88}},
  \bibinfo{pages}{063531} (\bibinfo{year}{2013}), \eprint{1306.0591}.

\bibitem[{\citenamefont{Urbanowski}(2015)}]{Urbanowski:2015ska}
\bibinfo{author}{\bibfnamefont{K.}~\bibnamefont{Urbanowski}}
  (\bibinfo{year}{2015}), \eprint{1502.03440}.

\bibitem[{\citenamefont{Lewis}()}]{CosmoMC}
\bibinfo{author}{\bibfnamefont{A.}~\bibnamefont{Lewis}},
  \emph{\bibinfo{title}{{CosmoMC}}},
  \bibinfo{howpublished}{\url{http://cosmologist.info/cosmomc/}}.

\bibitem[{\citenamefont{Lewis and Bridle}(2002)}]{Lewis:2002ah}
\bibinfo{author}{\bibfnamefont{A.}~\bibnamefont{Lewis}} \bibnamefont{and}
  \bibinfo{author}{\bibfnamefont{S.}~\bibnamefont{Bridle}},
  \bibinfo{journal}{Phys.Rev.} \textbf{\bibinfo{volume}{D66}},
  \bibinfo{pages}{103511} (\bibinfo{year}{2002}), \eprint{astro-ph/0205436}.

\bibitem[{\citenamefont{Feroz and Hobson}(2008)}]{Feroz:2007kg}
\bibinfo{author}{\bibfnamefont{F.}~\bibnamefont{Feroz}} \bibnamefont{and}
  \bibinfo{author}{\bibfnamefont{M.}~\bibnamefont{Hobson}},
  \bibinfo{journal}{Mon.Not.Roy.Astron.Soc.} \textbf{\bibinfo{volume}{384}},
  \bibinfo{pages}{449} (\bibinfo{year}{2008}), \eprint{0704.3704}.

\bibitem[{\citenamefont{Feroz et~al.}(2009)\citenamefont{Feroz, Hobson, and
  Bridges}}]{Feroz:2008xx}
\bibinfo{author}{\bibfnamefont{F.}~\bibnamefont{Feroz}},
  \bibinfo{author}{\bibfnamefont{M.}~\bibnamefont{Hobson}}, \bibnamefont{and}
  \bibinfo{author}{\bibfnamefont{M.}~\bibnamefont{Bridges}},
  \bibinfo{journal}{Mon.Not.Roy.Astron.Soc.} \textbf{\bibinfo{volume}{398}},
  \bibinfo{pages}{1601} (\bibinfo{year}{2009}), \eprint{0809.3437}.

\bibitem[{\citenamefont{Suzuki et~al.}(2012)\citenamefont{Suzuki, Rubin,
  Lidman, Aldering, Amanullah et~al.}}]{Suzuki:2011hu}
\bibinfo{author}{\bibfnamefont{N.}~\bibnamefont{Suzuki}},
  \bibinfo{author}{\bibfnamefont{D.}~\bibnamefont{Rubin}},
  \bibinfo{author}{\bibfnamefont{C.}~\bibnamefont{Lidman}},
  \bibinfo{author}{\bibfnamefont{G.}~\bibnamefont{Aldering}},
  \bibinfo{author}{\bibfnamefont{R.}~\bibnamefont{Amanullah}},
  \bibnamefont{et~al.}, \bibinfo{journal}{Astrophys.J.}
  \textbf{\bibinfo{volume}{746}}, \bibinfo{pages}{85} (\bibinfo{year}{2012}),
  \eprint{1105.3470}.

\bibitem[{\citenamefont{Chen et~al.}(2013)\citenamefont{Chen, Geng, Cao, Huang,
  and Zhu}}]{Chen:2013vea}
\bibinfo{author}{\bibfnamefont{Y.}~\bibnamefont{Chen}},
  \bibinfo{author}{\bibfnamefont{C.-Q.} \bibnamefont{Geng}},
  \bibinfo{author}{\bibfnamefont{S.}~\bibnamefont{Cao}},
  \bibinfo{author}{\bibfnamefont{Y.-M.} \bibnamefont{Huang}}, \bibnamefont{and}
  \bibinfo{author}{\bibfnamefont{Z.-H.} \bibnamefont{Zhu}}
  (\bibinfo{year}{2013}), \eprint{1312.1443}.

\bibitem[{\citenamefont{Eisenstein et~al.}(2005)}]{Eisenstein:2005su}
\bibinfo{author}{\bibfnamefont{D.~J.} \bibnamefont{Eisenstein}}
  \bibnamefont{et~al.} (\bibinfo{collaboration}{SDSS Collaboration}),
  \bibinfo{journal}{Astrophys.J.} \textbf{\bibinfo{volume}{633}},
  \bibinfo{pages}{560} (\bibinfo{year}{2005}), \eprint{astro-ph/0501171}.

\bibitem[{\citenamefont{Percival et~al.}(2010)}]{Percival:2009xn}
\bibinfo{author}{\bibfnamefont{W.~J.} \bibnamefont{Percival}}
  \bibnamefont{et~al.} (\bibinfo{collaboration}{SDSS Collaboration}),
  \bibinfo{journal}{Mon.Not.Roy.Astron.Soc.} \textbf{\bibinfo{volume}{401}},
  \bibinfo{pages}{2148} (\bibinfo{year}{2010}), \eprint{0907.1660}.

\bibitem[{\citenamefont{Eisenstein et~al.}(2011)}]{Eisenstein:2011sa}
\bibinfo{author}{\bibfnamefont{D.~J.} \bibnamefont{Eisenstein}}
  \bibnamefont{et~al.} (\bibinfo{collaboration}{SDSS Collaboration}),
  \bibinfo{journal}{Astron.J.} \textbf{\bibinfo{volume}{142}},
  \bibinfo{pages}{72} (\bibinfo{year}{2011}), \eprint{1101.1529}.

\bibitem[{\citenamefont{Ahn et~al.}(2014)}]{Ahn:2013gms}
\bibinfo{author}{\bibfnamefont{C.~P.} \bibnamefont{Ahn}} \bibnamefont{et~al.}
  (\bibinfo{collaboration}{SDSS Collaboration}),
  \bibinfo{journal}{Astrophys.J.Suppl.} \textbf{\bibinfo{volume}{211}},
  \bibinfo{pages}{17} (\bibinfo{year}{2014}), \eprint{1307.7735}.

\bibitem[{\citenamefont{Jones et~al.}(2009)\citenamefont{Jones, Read, Saunders,
  Colless, Jarrett et~al.}}]{Jones:2009yz}
\bibinfo{author}{\bibfnamefont{D.~H.} \bibnamefont{Jones}},
  \bibinfo{author}{\bibfnamefont{M.~A.} \bibnamefont{Read}},
  \bibinfo{author}{\bibfnamefont{W.}~\bibnamefont{Saunders}},
  \bibinfo{author}{\bibfnamefont{M.}~\bibnamefont{Colless}},
  \bibinfo{author}{\bibfnamefont{T.}~\bibnamefont{Jarrett}},
  \bibnamefont{et~al.}, \bibinfo{journal}{Mon.Not.Roy.Astron.Soc.}
  \textbf{\bibinfo{volume}{399}}, \bibinfo{pages}{683} (\bibinfo{year}{2009}),
  \eprint{0903.5451}.

\bibitem[{\citenamefont{Beutler et~al.}(2011)\citenamefont{Beutler, Blake,
  Colless, Jones, Staveley-Smith et~al.}}]{Beutler:2011hx}
\bibinfo{author}{\bibfnamefont{F.}~\bibnamefont{Beutler}},
  \bibinfo{author}{\bibfnamefont{C.}~\bibnamefont{Blake}},
  \bibinfo{author}{\bibfnamefont{M.}~\bibnamefont{Colless}},
  \bibinfo{author}{\bibfnamefont{D.~H.} \bibnamefont{Jones}},
  \bibinfo{author}{\bibfnamefont{L.}~\bibnamefont{Staveley-Smith}},
  \bibnamefont{et~al.}, \bibinfo{journal}{Mon.Not.Roy.Astron.Soc.}
  \textbf{\bibinfo{volume}{416}}, \bibinfo{pages}{3017} (\bibinfo{year}{2011}),
  \eprint{1106.3366}.

\bibitem[{\citenamefont{Drinkwater et~al.}(2010)\citenamefont{Drinkwater,
  Jurek, Blake, Woods, Pimbblet et~al.}}]{Drinkwater:2009sd}
\bibinfo{author}{\bibfnamefont{M.~J.} \bibnamefont{Drinkwater}},
  \bibinfo{author}{\bibfnamefont{R.~J.} \bibnamefont{Jurek}},
  \bibinfo{author}{\bibfnamefont{C.}~\bibnamefont{Blake}},
  \bibinfo{author}{\bibfnamefont{D.}~\bibnamefont{Woods}},
  \bibinfo{author}{\bibfnamefont{K.~A.} \bibnamefont{Pimbblet}},
  \bibnamefont{et~al.}, \bibinfo{journal}{Mon.Not.Roy.Astron.Soc.}
  \textbf{\bibinfo{volume}{401}}, \bibinfo{pages}{1429} (\bibinfo{year}{2010}),
  \eprint{0911.4246}.

\bibitem[{\citenamefont{Blake et~al.}(2011{\natexlab{a}})\citenamefont{Blake,
  Kazin, Beutler, Davis, Parkinson et~al.}}]{Blake:2011en}
\bibinfo{author}{\bibfnamefont{C.}~\bibnamefont{Blake}},
  \bibinfo{author}{\bibfnamefont{E.}~\bibnamefont{Kazin}},
  \bibinfo{author}{\bibfnamefont{F.}~\bibnamefont{Beutler}},
  \bibinfo{author}{\bibfnamefont{T.}~\bibnamefont{Davis}},
  \bibinfo{author}{\bibfnamefont{D.}~\bibnamefont{Parkinson}},
  \bibnamefont{et~al.}, \bibinfo{journal}{Mon.Not.Roy.Astron.Soc.}
  \textbf{\bibinfo{volume}{418}}, \bibinfo{pages}{1707}
  (\bibinfo{year}{2011}{\natexlab{a}}), \eprint{1108.2635}.

\bibitem[{\citenamefont{Blake et~al.}(2011{\natexlab{b}})\citenamefont{Blake,
  Davis, Poole, Parkinson, Brough et~al.}}]{Blake:2011wn}
\bibinfo{author}{\bibfnamefont{C.}~\bibnamefont{Blake}},
  \bibinfo{author}{\bibfnamefont{T.}~\bibnamefont{Davis}},
  \bibinfo{author}{\bibfnamefont{G.}~\bibnamefont{Poole}},
  \bibinfo{author}{\bibfnamefont{D.}~\bibnamefont{Parkinson}},
  \bibinfo{author}{\bibfnamefont{S.}~\bibnamefont{Brough}},
  \bibnamefont{et~al.}, \bibinfo{journal}{Mon.Not.Roy.Astron.Soc.}
  \textbf{\bibinfo{volume}{415}}, \bibinfo{pages}{2892}
  (\bibinfo{year}{2011}{\natexlab{b}}), \eprint{1105.2862}.

\bibitem[{\citenamefont{Alcock and Paczynski}(1979)}]{Alcock:1979mp}
\bibinfo{author}{\bibfnamefont{C.}~\bibnamefont{Alcock}} \bibnamefont{and}
  \bibinfo{author}{\bibfnamefont{B.}~\bibnamefont{Paczynski}},
  \bibinfo{journal}{Nature} \textbf{\bibinfo{volume}{281}},
  \bibinfo{pages}{358} (\bibinfo{year}{1979}).

\bibitem[{\citenamefont{Blake et~al.}(2011{\natexlab{c}})\citenamefont{Blake,
  Glazebrook, Davis, Brough, Colless et~al.}}]{Blake:2011ep}
\bibinfo{author}{\bibfnamefont{C.}~\bibnamefont{Blake}},
  \bibinfo{author}{\bibfnamefont{K.}~\bibnamefont{Glazebrook}},
  \bibinfo{author}{\bibfnamefont{T.}~\bibnamefont{Davis}},
  \bibinfo{author}{\bibfnamefont{S.}~\bibnamefont{Brough}},
  \bibinfo{author}{\bibfnamefont{M.}~\bibnamefont{Colless}},
  \bibnamefont{et~al.}, \bibinfo{journal}{Mon.Not.Roy.Astron.Soc.}
  \textbf{\bibinfo{volume}{418}}, \bibinfo{pages}{1725}
  (\bibinfo{year}{2011}{\natexlab{c}}), \eprint{1108.2637}.

\bibitem[{\citenamefont{Bond et~al.}(1997)\citenamefont{Bond, Efstathiou, and
  Tegmark}}]{Bond:1997wr}
\bibinfo{author}{\bibfnamefont{J.}~\bibnamefont{Bond}},
  \bibinfo{author}{\bibfnamefont{G.}~\bibnamefont{Efstathiou}},
  \bibnamefont{and} \bibinfo{author}{\bibfnamefont{M.}~\bibnamefont{Tegmark}},
  \bibinfo{journal}{Mon.Not.Roy.Astron.Soc.} \textbf{\bibinfo{volume}{291}},
  \bibinfo{pages}{L33} (\bibinfo{year}{1997}), \eprint{astro-ph/9702100}.

\bibitem[{\citenamefont{Li and Xia}(2013)}]{Li:2013awa}
\bibinfo{author}{\bibfnamefont{H.}~\bibnamefont{Li}} \bibnamefont{and}
  \bibinfo{author}{\bibfnamefont{J.-Q.} \bibnamefont{Xia}},
  \bibinfo{journal}{Phys.Lett.} \textbf{\bibinfo{volume}{B726}},
  \bibinfo{pages}{549} (\bibinfo{year}{2013}), \eprint{1309.0679}.

\bibitem[{\citenamefont{Liddle}(2004)}]{Liddle:2004nh}
\bibinfo{author}{\bibfnamefont{A.~R.} \bibnamefont{Liddle}},
  \bibinfo{journal}{Mon.Not.Roy.Astron.Soc.} \textbf{\bibinfo{volume}{351}},
  \bibinfo{pages}{L49} (\bibinfo{year}{2004}), \eprint{astro-ph/0401198}.

\bibitem[{\citenamefont{Mukherjee et~al.}(2006)\citenamefont{Mukherjee,
  Parkinson, Corasaniti, Liddle, and Kunz}}]{Mukherjee:2005tr}
\bibinfo{author}{\bibfnamefont{P.}~\bibnamefont{Mukherjee}},
  \bibinfo{author}{\bibfnamefont{D.}~\bibnamefont{Parkinson}},
  \bibinfo{author}{\bibfnamefont{P.~S.} \bibnamefont{Corasaniti}},
  \bibinfo{author}{\bibfnamefont{A.~R.} \bibnamefont{Liddle}},
  \bibnamefont{and} \bibinfo{author}{\bibfnamefont{M.}~\bibnamefont{Kunz}},
  \bibinfo{journal}{Mon.Not.Roy.Astron.Soc.} \textbf{\bibinfo{volume}{369}},
  \bibinfo{pages}{1725} (\bibinfo{year}{2006}), \eprint{astro-ph/0512484}.

\bibitem[{\citenamefont{Trotta}(2007{\natexlab{a}})}]{Trotta:2005ar}
\bibinfo{author}{\bibfnamefont{R.}~\bibnamefont{Trotta}},
  \bibinfo{journal}{Mon.Not.Roy.Astron.Soc.} \textbf{\bibinfo{volume}{378}},
  \bibinfo{pages}{72} (\bibinfo{year}{2007}{\natexlab{a}}),
  \eprint{astro-ph/0504022}.

\bibitem[{\citenamefont{Parkinson et~al.}(2005)\citenamefont{Parkinson,
  Tsujikawa, Bassett, and Amendola}}]{Parkinson:2004yx}
\bibinfo{author}{\bibfnamefont{D.}~\bibnamefont{Parkinson}},
  \bibinfo{author}{\bibfnamefont{S.}~\bibnamefont{Tsujikawa}},
  \bibinfo{author}{\bibfnamefont{B.~A.} \bibnamefont{Bassett}},
  \bibnamefont{and} \bibinfo{author}{\bibfnamefont{L.}~\bibnamefont{Amendola}},
  \bibinfo{journal}{Phys.Rev.} \textbf{\bibinfo{volume}{D71}},
  \bibinfo{pages}{063524} (\bibinfo{year}{2005}), \eprint{astro-ph/0409071}.

\bibitem[{\citenamefont{Liddle et~al.}(2006)\citenamefont{Liddle, Mukherjee,
  Parkinson, and Wang}}]{Liddle:2006kn}
\bibinfo{author}{\bibfnamefont{A.~R.} \bibnamefont{Liddle}},
  \bibinfo{author}{\bibfnamefont{P.}~\bibnamefont{Mukherjee}},
  \bibinfo{author}{\bibfnamefont{D.}~\bibnamefont{Parkinson}},
  \bibnamefont{and} \bibinfo{author}{\bibfnamefont{Y.}~\bibnamefont{Wang}},
  \bibinfo{journal}{Phys.Rev.} \textbf{\bibinfo{volume}{D74}},
  \bibinfo{pages}{123506} (\bibinfo{year}{2006}), \eprint{astro-ph/0610126}.

\bibitem[{\citenamefont{Parkinson et~al.}(2006)\citenamefont{Parkinson,
  Mukherjee, and Liddle}}]{Parkinson:2006ku}
\bibinfo{author}{\bibfnamefont{D.}~\bibnamefont{Parkinson}},
  \bibinfo{author}{\bibfnamefont{P.}~\bibnamefont{Mukherjee}},
  \bibnamefont{and} \bibinfo{author}{\bibfnamefont{A.~R.}
  \bibnamefont{Liddle}}, \bibinfo{journal}{Phys.Rev.}
  \textbf{\bibinfo{volume}{D73}}, \bibinfo{pages}{123523}
  (\bibinfo{year}{2006}), \eprint{astro-ph/0605003}.

\bibitem[{\citenamefont{Trotta}(2007{\natexlab{b}})}]{Trotta:2007hy}
\bibinfo{author}{\bibfnamefont{R.}~\bibnamefont{Trotta}},
  \bibinfo{journal}{Mon.Not.Roy.Astron.Soc.} \textbf{\bibinfo{volume}{378}},
  \bibinfo{pages}{819} (\bibinfo{year}{2007}{\natexlab{b}}),
  \eprint{astro-ph/0703063}.

\bibitem[{\citenamefont{Liddle}(2007)}]{Liddle:2007fy}
\bibinfo{author}{\bibfnamefont{A.~R.} \bibnamefont{Liddle}},
  \bibinfo{journal}{Mon.Not.Roy.Astron.Soc.} \textbf{\bibinfo{volume}{377}},
  \bibinfo{pages}{L74} (\bibinfo{year}{2007}), \eprint{astro-ph/0701113}.

\bibitem[{\citenamefont{Kurek and Szydlowski}(2008)}]{Kurek:2007tb}
\bibinfo{author}{\bibfnamefont{A.}~\bibnamefont{Kurek}} \bibnamefont{and}
  \bibinfo{author}{\bibfnamefont{M.}~\bibnamefont{Szydlowski}},
  \bibinfo{journal}{Astrophys.J.} \textbf{\bibinfo{volume}{675}},
  \bibinfo{pages}{1} (\bibinfo{year}{2008}), \eprint{astro-ph/0702484}.

\bibitem[{\citenamefont{Akaike}(1974)}]{Akaike:1974nl}
\bibinfo{author}{\bibfnamefont{H.}~\bibnamefont{Akaike}},
  \bibinfo{journal}{IEEE Trans. Auto. Control} \textbf{\bibinfo{volume}{19}},
  \bibinfo{pages}{716} (\bibinfo{year}{1974}).

\bibitem[{\citenamefont{Jeffreys}(1961)}]{Jeffreys:1961}
\bibinfo{author}{\bibfnamefont{H.}~\bibnamefont{Jeffreys}},
  \emph{\bibinfo{title}{Theory of Probability}} (\bibinfo{publisher}{Oxford
  University Press}, \bibinfo{address}{Oxford}, \bibinfo{year}{1961}),
  \bibinfo{edition}{3rd} ed.

\bibitem[{\citenamefont{MacKay}(2003)}]{MacKay:2003it}
\bibinfo{author}{\bibfnamefont{D.~J.~C.} \bibnamefont{MacKay}},
  \emph{\bibinfo{title}{Information Theory, Inference, and Learning
  Algorithms}} (\bibinfo{publisher}{Cambridge University Press},
  \bibinfo{address}{Cambridge}, \bibinfo{year}{2003}).

\bibitem[{\citenamefont{Schwarz}(1978)}]{Schwarz:1978ed}
\bibinfo{author}{\bibfnamefont{G.}~\bibnamefont{Schwarz}},
  \bibinfo{journal}{Annals of Statistics} \textbf{\bibinfo{volume}{6}},
  \bibinfo{pages}{461} (\bibinfo{year}{1978}).

\bibitem[{\citenamefont{Efstathiou}(2007)}]{Efstathiou:2007gz}
\bibinfo{author}{\bibfnamefont{G.}~\bibnamefont{Efstathiou}},
  \bibinfo{journal}{Nuovo Cim.} \textbf{\bibinfo{volume}{B122}},
  \bibinfo{pages}{1423} (\bibinfo{year}{2007}), \eprint{0712.1513}.

\bibitem[{\citenamefont{Chongchitnan and
  Efstathiou}(2007)}]{Chongchitnan:2007eb}
\bibinfo{author}{\bibfnamefont{S.}~\bibnamefont{Chongchitnan}}
  \bibnamefont{and}
  \bibinfo{author}{\bibfnamefont{G.}~\bibnamefont{Efstathiou}},
  \bibinfo{journal}{Phys.Rev.} \textbf{\bibinfo{volume}{D76}},
  \bibinfo{pages}{043508} (\bibinfo{year}{2007}), \eprint{0705.1955}.

\bibitem[{\citenamefont{Szydlowski et~al.}(2005)\citenamefont{Szydlowski,
  Krawiec, and Czaja}}]{Szydlowski:2004jv}
\bibinfo{author}{\bibfnamefont{M.}~\bibnamefont{Szydlowski}},
  \bibinfo{author}{\bibfnamefont{A.}~\bibnamefont{Krawiec}}, \bibnamefont{and}
  \bibinfo{author}{\bibfnamefont{W.}~\bibnamefont{Czaja}},
  \bibinfo{journal}{Phys. Rev.} \textbf{\bibinfo{volume}{E72}},
  \bibinfo{pages}{036221} (\bibinfo{year}{2005}), \eprint{astro-ph/0401293}.

\end{thebibliography}
\end{document}